\documentclass[final,6p,times,twocolumn,floatfix,prd,authoryear,nofootinbib,aps,eqsecnum]{revtex4-2}
\usepackage{bm}
\usepackage{newtxmath,newtxtext}
\usepackage{graphicx}
\usepackage{textcomp}
\usepackage{float}
\usepackage{booktabs}
\usepackage{dcolumn}
\usepackage{ragged2e}
\usepackage{hyperref}
\usepackage{xcolor}
\usepackage{epsfig}
\usepackage{appendix}
\usepackage{orcidlink}
\usepackage{multirow}
\usepackage{tcolorbox}
\usepackage[section]{placeins}
\makeatletter
\long\def\@makecaption#1#2{%
  \par
  \vskip\abovecaptionskip
  \begingroup
    \small\rmfamily
    \noindent
    \@make@capt@title{#1}{#2}\par
  \endgroup
  \vskip\belowcaptionskip
}
\makeatother
\hypersetup{colorlinks=true, linkcolor=red, urlcolor=blue, citecolor=red}
\newcommand{\bq}{\begin{equation}}
	\newcommand{\eq}{\end{equation}}
\newcommand{\bqn}{\begin{eqnarray}}
	\newcommand{\eqn}{\end{eqnarray}}
\newcommand{\nb}{\nonumber}
\newcommand{\lb}{\label}

\begin{document}
\title{High-Accuracy Semi-Analytic Mode Functions, Ultraviolet Regularity, and Particle Creation in the Hybrid Approach to mLQC-I}
\date{\today}

\author{Jamal Saeed}
\email{Jamal$\_$Saeed1@baylor.edu}

\author{Rui Pan}
\email{rui$\_$pan1@baylor.edu}

\author{Anzhong Wang \footnote{The corresponding author}}
\email{anzhong$\_$wang@baylor.edu; the corresponding author}
 
\affiliation{GCAP-CASPER, Department of Physics and Astronomy, Baylor University, Waco, TX 76798-7316, USA}

\begin{abstract}

We construct high-accuracy semi-analytic mode functions for scalar cosmological perturbations in mLQC-I within the hybrid approach and use them to study ultraviolet regularity and gravitational particle creation. The evolution naturally separates into an asymptotic pre-de Sitter branch, a pre-inflationary region containing the quantum bounce, and the subsequent inflationary phase. In the intermediate region, we extend the uniform asymptotic approximation beyond the leading order, construct the complete third-order solution, and introduce an adaptive defect correction. The third-order approximation achieves sub-percent complex-mode accuracy for representative higher-wavenumber modes, whereas the defect-corrected solutions agree with direct numerical evolution at the \({\cal O}(10^{-5}\%)\) level over the tested low-wavenumber range. The resulting construction retains both amplitudes and phases and provides a controlled basis for future precision calculations of the primordial power spectrum and non-Gaussian correlation functions. We then compare two initial-state prescriptions: the de Sitter vacuum selected in the remote contracting branch and a fourth-order adiabatic state imposed at the bounce. A central result is that finite-wavenumber matching and the strict ultraviolet limit require different asymptotic treatments. Extrapolating the leading connection to infinite wavenumber produces a spurious constant negative-frequency coefficient and an apparent ultraviolet divergence. Instead, the large-wavenumber de Sitter mode agrees with the fourth-order adiabatic expansion through the order required for energy-momentum tensor renormalization, and this ultraviolet structure is preserved by smooth exact evolution. However, finite and intermediate modes can undergo nonadiabatic Bogoliubov mixing and particle creation. Thus, finite-mode particle production and ultraviolet regularity coexist naturally. The ultraviolet argument also extends to the dressed-metric approach in its corresponding large-wavenumber regime, although the quantitative mode functions for any given wavenumber must be recalculated using the dressed-metric effective mass.

\end{abstract}

\maketitle

\section{Introduction}
\lb{Sec:introduction}

Inflation provides a remarkably successful framework for describing the
early universe. In addition to resolving several conceptual problems of
standard big-bang cosmology, it provides a causal mechanism for generating
primordial perturbations that subsequently seed the large-scale structure of
the universe and the anisotropies of the cosmic microwave background (CMB)
\cite{Planck:2018jri}. Nevertheless, if the inflationary phase is extended
sufficiently far into the past, observable modes can be traced back to
physical wavelengths at which quantum-gravitational effects are expected to
become important. Moreover, within general relativity, the inflationary
spacetime is past incomplete, and the presence of the big-bang singularity
prevents one from following the cosmological evolution, and hence the
perturbations, arbitrarily far into the past. These considerations provide an
important motivation for studying cosmological perturbations in theories of
quantum gravity.

Loop quantum cosmology (LQC) provides one such framework, in which the
classical big-bang singularity is generically replaced by a quantum bounce
due to quantum-geometric effects
\cite{Ashtekar:2011ni,Agullo:2023rqq}. As a result, the pre-inflationary
evolution can be extended through the Planck regime into a contracting branch,
making it possible, at least in principle, to follow cosmological
perturbations from the remote contracting universe through the quantum bounce
and into the slow-roll inflationary phase. Several approaches have been
developed to describe cosmological perturbations in LQC. Among them, the
dressed metric approach
\cite{Agullo:2012sh,Agullo:2012fc,Agullo:2013ai} and the hybrid approach
\cite{Fernandez-Mendez:2012poe,Fernandez-Mendez:2013jqa,
Fernandez-Mendez:2014raa,ElizagaNavascues:2020uyf}
have been extensively investigated. Although they have different origins
and lead to different effective mass functions in the Planck regime, their
predictions are closely related in appropriate limits
\cite{Li:2022evi,Iteanu:2022zha}.

In the present work, we focus on a modified version of LQC, usually referred
to as mLQC-I, in which the Euclidean and Lorentzian parts of the
gravitational Hamiltonian are treated separately in the quantization
procedure
\cite{Yang:2009fp,Assanioussi:2018hee,Li:2018opr} \footnote{In \cite{Dapor:2017rwv,Han:2021cwb}, the authors adopted a different approach (the so-called top-down approach), and a similar Hamiltonian was obtained.}.
While the big-bang singularity is again replaced by a quantum bounce, the
resulting background evolution is intrinsically asymmetric with respect to
the bounce. In particular, the post-bounce branch approaches the classical
general-relativistic evolution, whereas the remote contracting branch
approaches an effective de Sitter spacetime
\cite{Li:2018opr,Li:2018fco,Li:2019ipm,Saeed:2024xhk}.
For kinetic-energy-dominated initial conditions at the bounce, the background
evolution exhibits remarkable universal properties, and slow-roll inflation
is generated generically
\cite{Li:2019ipm}
\footnote{Recently, in the Bianchi~I model, a quantum mechanism of shear
damping was proposed, which can lead to a complementary branch structure.
In this scenario, the pre-bounce branch is asymptotically described by
classical general relativity, whereas the post-bounce branch approaches a
quantum inflationary phase characterized by an effective Planck-scale
cosmological constant \cite{Gan:2025uvt,Gan:2026ivx}.}.
Cosmological perturbations in mLQC-I have been studied in both the dressed
metric and hybrid approaches
\cite{Agullo:2018wbf,Li:2019qzr,Li:2020mfi,Pan:2025ppa}.

A basic ingredient in extracting predictions from these models is an accurate
determination of the perturbative mode functions, so theoretical predictions are 
accurate enough in order to compare with current and forthcoming observations \cite{Planck:2018jri,AtacamaCosmologyTelescope:2025blo,AtacamaCosmologyTelescope:2025nti,CMB-S4:2016ple}. The primordial power
spectrum depends directly on their late-time amplitudes, while higher-order
correlation functions, including those characterizing primordial
non-Gaussianity, are also sensitive to their relative phases and to their
time dependence through the interaction integrals. Thus, for precision
studies of primordial observables, it is useful to construct high-accuracy
semi-analytic mode functions that remain reliable across the
pre-inflationary dynamics, including the quantum-bounce regime. This is
particularly nontrivial in mLQC-I, where the effective mass changes sign and
the mode equation develops a turning point before the onset of inflation.

A second important issue is the choice of the quantum state of the
perturbations. In a generic time-dependent curved spacetime, there is no
unique notion of a vacuum, and different choices may lead to physically
different predictions. Closely related to this ambiguity is gravitational
particle creation: the time dependence of the background can mix positive-
and negative-frequency modes, producing particles even when the field is
initially placed in a vacuum state
\cite{Parker:1969au,Parker_Toms_2009}. The ultraviolet behavior of the
corresponding energy-momentum tensor must then be treated with an appropriate
renormalization prescription. For spatially homogeneous cosmological
spacetimes, adiabatic regularization provides a particularly useful framework
\cite{PhysRevD.9.341,Parker_Toms_2009,AgulloNavarroSalasOlmoParker2010}.
The ultraviolet behavior of the
renormalized energy density therefore provides an important consistency test
of a proposed perturbative state.

A number of prescriptions for selecting the initial state have been
considered in LQC. In the dressed-metric approach, fourth-order adiabatic
states imposed at the quantum bounce have played an important role
\cite{Agullo:2012sh,Agullo:2012fc,Agullo:2013ai,AgulloNelsonAshtekar2015}. 
Other proposals include a
quantum generalization of Penrose's Weyl-curvature hypothesis due to Ashtekar
and Gupt \cite{Ashtekar:2016wpi,Ashtekar:2016pqn}, prescriptions based on
minimizing oscillations in the primordial power spectra
\cite{deBlas:2016puz,CastelloGomar:2017kbo,ElizagaNavascues:2020fai,Giesel:2026fsp},
and states selected in the contracting branch \cite{Zhu:2017jew,Agullo:2017eyh,AgulloMorris2015,AgulloOlmedoSreenath2020PRL,AgulloOlmedoSreenath2020PRD}.
In mLQC-I, the latter
possibility is especially attractive because the remote contracting
spacetime becomes asymptotically de Sitter. Since de Sitter spacetime is
maximally symmetric, a vacuum state respecting these symmetries provides a
particularly natural choice \cite{Agullo:2018wbf}.
In our recent work \cite{Pan:2025ppa}, we
showed that the corresponding de Sitter state in the remote contracting phase
is stable and minimizes particle creation, even when some of the modes are
not in an adiabatic regime \cite{Penna-Lima:2022dmx}. In addition, the
asymptotic Hamiltonian diagonalization prescription leads to the same initial
condition \cite{ElizagaNavascues:2019itm}. Thus, from the viewpoint of the
asymptotic background itself, this state possesses several desirable physical
properties.

The subsequent evolution, however, raises two related questions. First, how
accurately can the mode functions be represented and propagated through the
strongly time-dependent pre-inflationary regime? Second, even if an initial
state is well motivated where it is specified, how does the subsequent
nonadiabatic evolution affect particle creation and the ultraviolet behavior
of the state? Particle creation in LQC has been investigated in several
settings. In particular, the pre-inflationary dynamics and the associated
Bogoliubov mixing were studied for adiabatic states and for states specified
in the contracting branch
\cite{Agullo:2012sh,Agullo:2012fc,Agullo:2013ai,Zhu:2017jew,Agullo:2017eyh}.
The gravitational production of particles during the bounce and its possible
backreaction were subsequently analyzed in
Refs.~\cite{Graef:2020gpp,Vicente:2022gpp}. These studies demonstrate that
particle production can provide a sensitive test of the validity of the
test-field approximation and of the perturbative description itself.

The present paper addresses these questions in the hybrid approach to
mLQC-I. Our first goal is to construct high-accuracy semi-analytic complex
mode functions from the remote contracting branch, through the quantum
bounce, and into the inflationary phase. The behavior of the hybrid effective
mass naturally separates the evolution into the pre-de Sitter,
pre-inflationary, and inflationary regimes. The mode functions are known
analytically in the two asymptotic regimes. In the intermediate region, which
contains the quantum bounce, the modified Mukhanov--Sasaki equation has a
single simple turning point for the modes considered here. We therefore
employ the uniform asymptotic approximation (UAA)
\cite{Olver:1997,Zhu:2013upa,Zhu:2014aea,Zhu:2016srz,Pan:2023aiv}.
We go beyond the leading approximation by constructing the complete
third-order UAA, analyzing the next error-control coefficient, and
introducing an adaptive defect correction for the modes for which the
ordinary hierarchy converges slowly. The resulting construction retains
the analytic organization of the UAA while providing high-accuracy
semi-analytic mode functions. We also formulate the finite-wavenumber
connection between the remote contracting and inflationary regimes in
transfer-matrix form, preserving both amplitudes and relative phases. These
results provide a controlled starting point for future precision calculations
of the primordial power spectrum and non-Gaussian correlation functions.

Our second goal is to study the regularity of ultraviolet radiation and the creation of particles for
two physically motivated initial-state prescriptions. The first is the de
Sitter vacuum selected in the remote contracting phase
\cite{Li:2019qzr,Pan:2025ppa}, while the second is a fourth-order adiabatic
state imposed directly at the quantum bounce
\cite{Agullo:2012sh,Agullo:2012fc,Agullo:2013ai}. Comparing these two
prescriptions allows us to distinguish properties associated with the choice
of the initial state from those generated by its subsequent propagation
through a nonadiabatic background.

A central point of our analysis is that the finite-wavenumber
turning-point problem and the strict ultraviolet limit require different
asymptotic treatments. The UAA connection formulas are derived in a
small overlap region $y$ with $y=k\eta\ll1$, and provide an accurate
description of finite and intermediate modes. However, at fixed nonzero conformal
time, the strict ultraviolet limit $k\rightarrow\infty$ implies
$y\rightarrow\infty$. These two limits cannot be interchanged. In
particular, extrapolating the leading  connection directly to
$k\rightarrow\infty$ produces a $k$-independent negative-frequency
coefficient and an apparent quartic divergence after fourth-order adiabatic
subtraction. We show that this behavior is an artifact of applying a
finite-wavenumber connection formula outside the asymptotic regime required
to determine the ultraviolet Bogoliubov coefficient.

For the  de Sitter vacuum, the ultraviolet sector must instead be
analyzed directly using the large-$k$ Hankel expansion and the fourth-order
adiabatic expansion. The two agree through the order required for
energy-momentum tensor renormalization, and the corresponding ultraviolet adiabatic
structure is preserved by exact evolution for the smooth effective mass
considered here. Nonadiabatic evolution can nevertheless mix positive- and
negative-frequency components for finite and intermediate wavenumbers and
produce particles. The fourth-order adiabatic state imposed at the bounce is
likewise ultraviolet regular, while its subsequent evolution can generate a
nonzero finite-wavenumber particle population. Thus, finite-mode particle
creation and ultraviolet regularity are complementary rather than mutually
exclusive properties of the evolving perturbative state.

The quantitative mode-function analysis of Secs.~II and III is specific to
the hybrid approach because it depends on the corresponding effective mass
and turning-point structure. The ultraviolet-regularity argument and the
framework for particle creation, however, are more general. In particular,
in the large-$k$ regime relevant to the ultraviolet analysis, the
dressed-metric mode equation likewise possesses a single simple turning point
and admits an Airy-type uniform approximation \cite{Pan:2025ppa}. The same
distinction between finite-$k$ Airy matching and the strict large-$k$
adiabatic limit therefore applies there as well, although the quantitative
mode functions for any given wavenumber and the corresponding particle
spectrum must be recalculated using the dressed-metric effective mass.

The remainder of this paper is organized as follows. In Sec.~II, we review
the modified Mukhanov--Sasaki (MS) equation for scalar perturbations in the hybrid
approach to mLQC-I and analyze the corresponding effective mass function. In
Sec.~III, we construct the mode functions in the pre-inflationary phase,
develop the complete third-order UAA and its error control, introduce the
adaptive defect correction, and derive the finite-wavenumber transfer
matrices connecting the remote contracting and inflationary regimes. In
Sec.~IV, we impose the de Sitter vacuum in the remote contracting phase and
study finite-mode particle creation and ultraviolet regularity by separating
the finite-$k$ matching problem from the strict large-$k$ adiabatic sector.
In Sec.~V, we analyze the same questions for a fourth-order adiabatic state
imposed directly at the quantum bounce. Section~VI summarizes our main
results and discusses future applications to primordial observables and the
dressed-metric approach. Appendix~A presents a compact fundamental-matrix
representation of the defect correction introduced in Secs.~III D and III E,
together with the corresponding Wronskian consistency condition for the
corrected transfer matrix.

In the following, we use units in which $\hbar=c=1$ and keep Newton's
constant $G$ explicit through $\kappa=8\pi G$. We use $t_P$ and $m_P$ to
denote the Planck time and Planck mass, respectively. Unless otherwise stated,
all numerical results and figures are presented in units in which
$t_P=m_P=1$.

\section{Cosmological Perturbations in the Hybrid Approach and the Modified MS Equation}
\lb{sec:effective_mass_function}

In this section, we review the modified MS equation for
scalar perturbations in mLQC-I within the hybrid approach
\cite{Li:2020mfi,ElizagaNavascues:2020uyf}. Tensor perturbations can be
treated analogously, with the corresponding tensor effective mass. The
modified MS equation for scalar perturbations takes the form  
\bq
\lb{MS_equation}
\nu^{\prime\prime}_{k}(\eta) +\left(k^2+m_{\text{eff}}^2\right)\nu_{k}(\eta)=0,
\eq
where $\nu_k(\eta)$ is the MS variable, related to the comoving curvature
perturbation $\mathcal{R}_k$ by $\nu_k=z_s\mathcal{R}_k$, with
$z_s=a\dot{\phi}/H$. A prime denotes differentiation with respect to
conformal time $\eta$, while an overdot denotes differentiation with respect
to cosmic time $t$, where $d\eta=dt/a(t)$. The constant $k$ denotes the
comoving wavenumber. The observationally relevant range of modes can be
written as \cite{Ashtekar:2011ni} 
\bq
\frac{k}{a_0}\in (0.1,10^3)\frac{k_*}{a_0},
\lb{eq:II-window-physical}
\eq
where $k_*/a_0\equiv0.05\,{\rm Mpc}^{-1}$ denotes the physical pivot
wavenumber today used in the Planck analysis \cite{Planck:2018jri},
while $k_*$ is the corresponding comoving pivot wavenumber. Here
$a_0\equiv e^{N_{\rm tot}}a_B$ is the present value of the scale factor,
and $N_{\rm tot}$ denotes the total number of e-folds of expansion from
the quantum bounce to the present epoch. Thus, for
$N_{\rm tot}\simeq141$ \cite{Ashtekar:2011ni,Zhu:2017jew}, and with
$a_B=1$, the corresponding range of comoving wavenumbers normalized at
the bounce is
\bqn
k\in(0.1,10^3)k_*
&=&(0.1,10^3)a_0\left(0.05\,{\rm Mpc}^{-1}\right)\nb\\
&\simeq& (1,10^4)m_P.
\lb{eq:II-window-comoving}
\eqn
 The mode function $\nu_k(\eta)$ is normalized by the Wronskian condition
\bq
\lb{eq3.2}
\nu^{*}_{k}\nu^{\prime}_{k} - \nu^{*\prime }_{k}\nu_{k} = -i.
\eq
In terms of the mode function $\nu_k(\eta)$, the primordial scalar power
spectrum is given by 
\bq
\lb{scalarpower}
P_\mathcal R=\frac{k^3}{2\pi^2}\frac{|\nu_{ k}|^2}{z^2_s}.
\eq
The specific form of the mass term in the MS equation depends on the
approach used to describe cosmological perturbations
\cite{Agullo:2023rqq,ElizagaNavascues:2020uyf}. In the hybrid approach,
the classical mass function that serves as the starting point for the
polymerization is \cite{Li:2020mfi}
\begin{widetext}
\bq
\lb{eq3.4}
m_{\text{hybrid}}^2 =\frac{\kappa p^2_\phi}{6v^{4/3}}\left(19- \frac{3\kappa\gamma^2 p^2_\phi}{\Omega^2}\right)+v^{2/3}\left(V_{, \phi\phi}+\frac{2\kappa\gamma p_\phi }{\Omega}V_{,\phi} - \frac{2\kappa}{3}V\right),
\eq
\end{widetext}
where $\Omega=\gamma a^3H$, while $p_\phi$ and $V(\phi)$ denote,
respectively, the momentum conjugate to the scalar field $\phi$ and its
potential. We use $V_{,\phi}\equiv dV/d\phi$,
$V_{,\phi\phi}\equiv d^2V/d\phi^2$, and analogous notation for higher
derivatives.

The effective hybrid mass is obtained by polymerizing the classical
expression in Eq.~(2.6). In particular, the terms involving
$\pi_a^{-1}$ and $\pi_a^{-2}$ require special care, since classically
\cite{Ashtekar:2011ni}
\bq
\lb{eq3.4a}
\pi_a = -\frac{3a\dot a}{4\pi G},
\eq
and hence the classical expression for $\pi_a$ vanishes at the quantum
bounce, where $H\equiv\dot a/a=0$. Consequently, the inverse powers of
$\pi_a$ appearing in the classical mass function cannot be used directly
through the bounce and must be replaced by expressions that admit a
well-defined continuation to the pre-bounce branch. This continuation must
also be consistent with the polymerization of the background dynamics.
Recall that the classical Hamiltonian constraint for the homogeneous
background takes the form 
\bqn
\mathcal H^{(0)}=-\frac{3\Omega^2}{8\pi G v\gamma^2}+\frac{p^2_\phi}{2v}+v V. 
\eqn 
In mLQC-I, the corresponding effective Hamiltonian constraint is obtained by
polymerizing the classical expression through the replacement
\cite{Li:2020mfi}
\bqn
 \Omega^2\rightarrow\Omega^2_{{\scriptscriptstyle{\mathrm{I}}}} &=& -\frac{v^2\gamma^2}{\lambda^2}\Big\{\sin^2\left(\lambda b\right)
 -\frac{\gamma^2+1}{4\gamma^2} \sin^2\left(2 \lambda b\right) \Big\}. ~~~~~
\eqn
Guided by this background polymerization, we adopt the following
prescription for the inverse powers of $\Omega$ appearing in
Eq.~(2.6) \cite{Li:2020mfi}:
\bqn
&& \frac{1}{\Omega^2}\rightarrow \frac{1}{\Omega^2_{{\scriptscriptstyle{\mathrm{I}}}}},  \;\;\;  \frac{1}{\Omega}\rightarrow \frac{\Lambda_{{\scriptscriptstyle{\mathrm{I}}}}}{\Omega^2_{{\scriptscriptstyle{\mathrm{I}}}}},\;\;\;
\Lambda_{{\scriptscriptstyle{\mathrm{I}}}} \equiv \frac{v}{2\lambda}\sin\left(2\lambda b\right). ~~~
\eqn 
The first replacement ensures consistency with the polymerization of the
background Hamiltonian, while the second is chosen to respect the
superselection structure of the background dynamics.\footnote{We refer the
reader to Ref.~\cite{Li:2020mfi} for further details of the polymerization
prescription.} With these replacements, the effective hybrid mass entering
the modified MS equation becomes 
\bqn
\lb{eq3.9}
m^2_{\mathrm{eff}} &=& \frac{4 \pi G\, p_{\phi}^2}{3\, v^{4/3}} \left(19 - \frac{24 \pi G\, \gamma^2\, p_{\phi}^2}{\Omega_{I}^2} \right) \nb\\
&& 
+ v^{2/3} \left( 
V_{,\phi\phi} 
+ \frac{16 \pi G\, \gamma\, p_{\phi}\, \Lambda_{I}}{\Omega_{I}^2} V_{,\phi}\right.\nb\\
&& ~~~~~~~~~~~~~~~ \left. - \frac{16 \pi G}{3} V \right).
\eqn


\begin{figure}[htp]
{
\includegraphics[width=8cm]{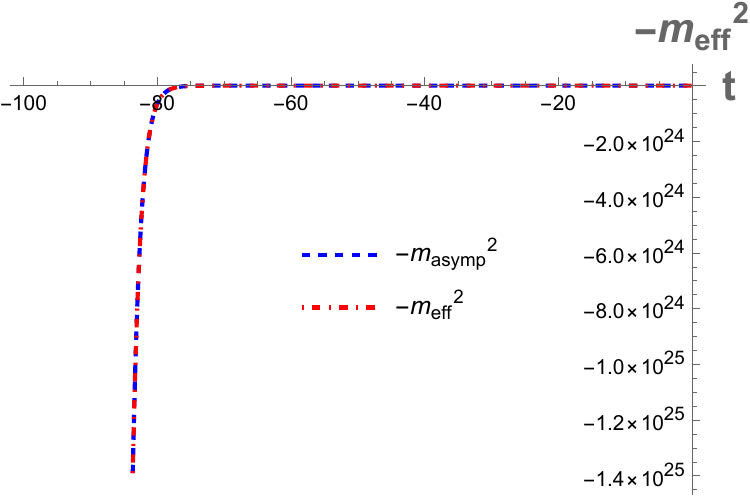}\\
(a)\\
\includegraphics[width=8cm]{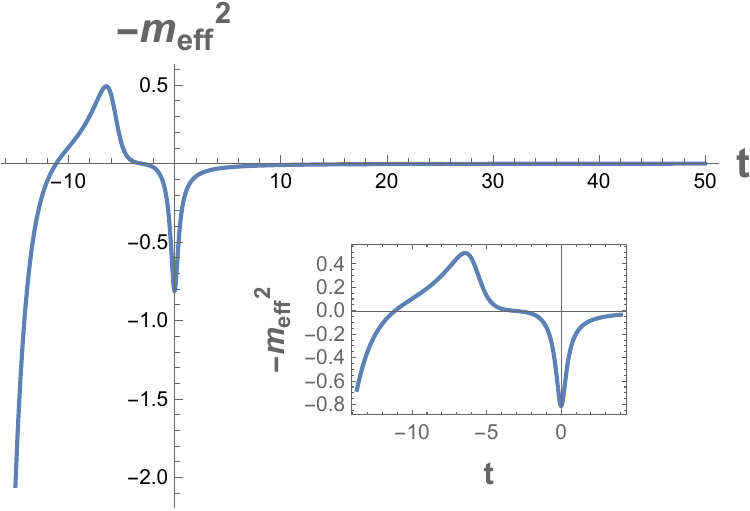}\\
(b)\\
\includegraphics[width=8cm]{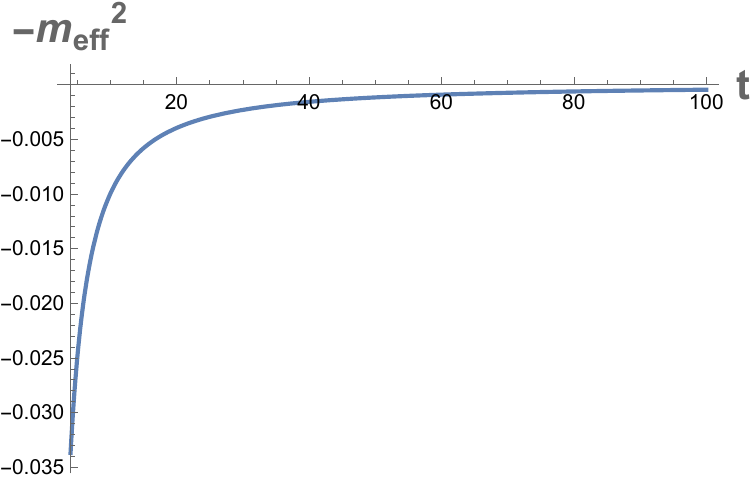}\\
(c)\\
}
\caption{The effective mass function defined by Eq.~(\ref{eq3.9}) over different time intervals: (a) $-100<t/t_P<0$; (b) $-15<t/t_P<50$; and (c) $0<t/t_P<100$.}
\label{fig1}
\end{figure}


\begin{figure}[t!]
{ 
\includegraphics[width=8cm]{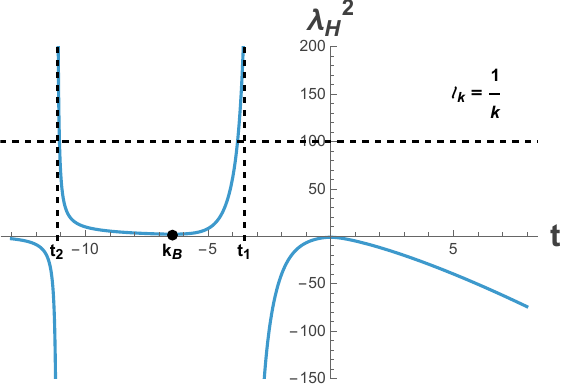} 
}
\caption{The effective-horizon quantity $\lambda_H^2$ defined by
Eq.~(\ref{eq3.15}).}  
\label{fig2}
\end{figure}

Figure~\ref{fig1} shows the evolution of
$\omega^2(t)\equiv -m_{\rm eff}^2(t)$ over several time intervals,
including its asymptotic behavior as $t\rightarrow\pm\infty$.
In particular, the effective mass changes sign according to 
\bqn
\lb{eq3.14aa}
m_{\text{eff}}^2 = \begin{cases}
    > 0, & t > t_1, \cr
    < 0, & t_2 < t < t_1, \cr
    > 0, & t < t_2, \cr
\end{cases} 
\eqn
where $t_1\simeq-3.22\,t_P$ and $t_2\simeq-11.15\,t_P$. Within the interval $t_2<t<t_1$, $\omega^2(t)$ reaches a positive
maximum. It is convenient to characterize this maximum by the
wavenumber scale $K_\gamma$, defined through
\bqn
\lb{eq3.14bb}
K_\gamma\equiv\sqrt{\max_t\omega^2(t)}
\simeq0.4916\,m_P.
\eqn
 To characterize the mode evolution, we introduce the effective-horizon
quantity $\lambda_H^2$ by 
\bqn
\lb{eq3.15}
\lambda_H^2 \equiv  \frac{1}{\omega^2(t)}.  
\eqn
In terms of this quantity, the MS equation~(\ref{MS_equation}) can be
written as 
\bqn
\lb{eq3.15aa}
\nu^{\prime\prime}_{k}(\eta) + \omega_k^2(\eta)\nu_{k}(\eta)=0,
\eqn
where the effective frequency is
\bqn
\lb{MS_equationB} 
\omega_k^2(\eta) &\equiv& k^2-\omega^2(t(\eta))
= \frac{1}{\ell_k^2}-\frac{1}{\lambda_H^2},
\quad \ell_k\equiv k^{-1}. ~~~
\eqn
When $\lambda_H^2>0$, Eq.~(\ref{MS_equationB}) gives
$\omega_k^2>0$ for $\ell_k<\lambda_H$ and $\omega_k^2<0$ for
$\ell_k>\lambda_H$. In the asymptotic de Sitter regime,
$\lambda_H$ reduces to the comoving Hubble radius,
$\lambda_H=L_H\equiv1/(aH)$. In general, however,
$\lambda_H^2$ can become negative. In that case no real effective-horizon
scale exists, and Eq.~(\ref{MS_equationB}) gives $\omega_k^2>0$ for all
real $k$; consequently, the usual inside/outside-horizon terminology is
not applicable. 

We emphasize that $\omega_k^2>0$ by itself does not imply adiabatic
evolution. For sufficiently large $k$, the frequency is dominated by
$k^2$ and the ultraviolet modes evolve adiabatically, whereas the
adiabaticity of finite-wavenumber modes must be assessed from the time
variation of $\omega_k$, as discussed below. This distinction is
particularly important when imposing initial conditions in the remote
contracting phase, $t_i\ll t_2$.

As illustrated in Fig.~\ref{fig2}, the evolution of the mode functions can
be divided into three regimes: (i) the pre-de Sitter phase,
$-\infty<t\lesssim t_2$; (ii) the pre-inflationary phase,
$t_2\lesssim t\lesssim t_{\rm inf}$; and (iii) the inflationary phase,
$t_{\rm inf}\lesssim t\lesssim t_{\rm end}$, where $t_{\rm inf}$ and
$t_{\rm end}$ denote the onset and end of inflation, respectively.
In the remainder of this section, we consider the mode functions in the
pre-de Sitter and inflationary phases. The intermediate pre-inflationary
phase, including the quantum bounce, is studied in Sec.~III.

 \subsection{Mode Function in the Pre-de Sitter Phase}

In the asymptotic remote contracting limit, $t\rightarrow-\infty\; (\eta \rightarrow 0)$, the
background behaves as 
\bqn
\lb{eq3.9aa}
a(t) &\simeq& a_me^{-H_{\Lambda}(t - t_m)} = \frac{1}{H_{\Lambda} \eta}, \nb\\
\phi (t) &\simeq&   \phi_{\infty}, \nb\\
p_{\phi} &\simeq& \frac{V_{,\phi}}{3H_{\Lambda}}a^3,\quad  b(t)  \simeq b_{\infty}, 
\eqn
where $\phi_\infty$ and $b_\infty$ are constants characterizing the
asymptotic pre-de Sitter branch. Substituting these asymptotic expressions
into Eq.~(\ref{eq3.9}), the effective mass reduces to
\bq
\small
\lb{eq3.10}
m^2_{\mathrm{eff}} \simeq a^2\left(t\right)\epsilon_{\lambda} = \frac{\epsilon_{\Lambda}}{\eta^2},
\eq
where
\bqn
\lb{eq3.10aa}
\epsilon_{\Lambda} &\equiv& \frac{\epsilon_{\lambda}}{H_{\Lambda}^2} 
= \frac{\kappa}{54\,H_{\Lambda}^4}\left(19 - \frac{\kappa\lambda^2V_{,\phi}^2}{3H_{\Lambda}^2{\cal{D}}(b_{\infty})}\right)V_{,\phi}^2\nb\\ 
&& - \frac{1}{H_{\Lambda}^2}\Bigg[\frac{2\kappa}{3}V 
 -   \frac{\kappa\lambda\sin(2\lambda b_{\infty})}{3\gamma H_{\Lambda} {\cal{D}}(b_{\infty})}V_{,\phi}^2  -  V_{,\phi\phi}\Bigg], 
\eqn
where $V\equiv V(\phi_\infty)$, and all derivatives of the potential,
such as $V_{,\phi}$ and $V_{,\phi\phi}$, are evaluated at
$\phi=\phi_\infty$. We have also defined
\bqn
\lb{eq3.10cc}
{\cal{D}}(b_{\infty}) &\equiv& \frac{\gamma^2+1}{4\gamma^2}\sin^2(2\lambda b_{\infty}) - \sin^2(\lambda b_{\infty}). 
\eqn
For the background considered here, one finds
$\epsilon_\Lambda\simeq{\cal O}(10^{-5})$. Substituting Eq.~(\ref{eq3.10})
into the MS equation~(\ref{MS_equation}) gives
\bqn
\lb{sol3B}
\nu_k &\simeq& y^{1/2}\left[a_k H_{\mu}^{(2)}(y) + b_k H_{\mu}^{(1)}(y)\right], \nb\\
  \mu &\equiv& \left(\frac{1}{4} - \epsilon_{\Lambda}\right)^{1/2}, ~~~
\eqn
where $a_k$ and $b_k$ are integration constants,
$y\equiv k\eta$, and $H_\mu^{(1)}(y)$ and $H_\mu^{(2)}(y)$ denote
the Hankel functions of the first and second kinds, respectively.
 
Since $\epsilon_\Lambda\simeq{\cal O}(10^{-5})$, one has
$\mu=1/2+{\cal O}(\epsilon_\Lambda)$. For the leading small-$y$
behavior displayed below, we may therefore set $\mu=1/2$. The full
$\mu$ dependence will be retained later when deriving the
finite-wavenumber matching conditions. As $y\rightarrow0^+$, the
corresponding Hankel functions behave as
\bqn
\lb{eq4.10aa}
H^{(1)}_{1/2}(y)  = H^{*\; (2)}_{1/2}(y) \simeq \sqrt{\frac{2}{\pi y}}\, e^{i(y - \frac{\pi}{2})}, \; ( y \rightarrow 0^+), ~~~
\eqn
  and hence
\bqn 
\lb{eq4.11}
\nu_k(\eta) &\simeq& \sqrt{\frac{2}{\pi}} \Big[\left(a_k + b_k\right)y    + i  \left(a_k - b_k\right)\Big] + {\cal{O}}\left(y^2\right). ~~~~~~~~~~~
\eqn

 \subsection{Mode Functions in the Inflationary Phase}

 During the inflationary phase, once a mode enters the oscillatory regime
$k^2\gg |m_{\rm eff}^2|$, the effective frequency approaches $k$, and
the MS mode takes the leading plane-wave form  
\bq
\nu_k(\eta)=\frac{1}{\sqrt{2k}}
\left(\alpha_k e^{-iy}+\beta_k e^{iy}\right),
\quad k^2\gg |m_{\rm eff}^2|,
\lb{sol3A}
\eq
where $\alpha_k$ and $\beta_k$ are the positive- and negative-frequency
coefficients in this oscillatory basis. Their values are fixed by the
initial state and by the evolution through the pre-inflationary region.

In the next section, we use the UAA developed in
Refs.~\cite{Olver:1997,Zhu:2013upa,Zhu:2014aea,Zhu:2016srz,Pan:2023aiv}
to construct the mode functions throughout the pre-inflationary region,
including the quantum bounce. The resulting solution is then matched to
the asymptotic pre-de Sitter solution~(\ref{sol3B}) and to the oscillatory
inflationary solution~(\ref{sol3A}), providing a global description of the
mode evolution from the remote contracting branch through the bounce and
into the inflationary phase.

\FloatBarrier
\section{Mode Functions in the Pre-Inflationary Phase and Matching Conditions}
\lb{sec:Analytical_Solutions}

In this section, we use the UAA to solve the MS equation
(\ref{MS_equation}) throughout the pre-inflationary phase,
$t_2\lesssim t\lesssim t_{\rm inf}$, and then connect this solution
to the pre-de Sitter and inflationary solutions obtained in Sec.~II.
We first construct the one-turning-point Airy approximation, then extend
it to the complete third order and quantify its accuracy by comparison
with direct numerical evolution. We subsequently examine the next UAA
error-control coefficient and introduce a defect-corrected approximation
for the modes for which the ordinary Airy hierarchy converges slowly.

\subsection{Mode Functions in the Pre-Inflationary Phase}

Using $y\equiv k\eta$, the MS equation can be written in the standard
form appropriate for the uniform asymptotic approximation
\cite{Olver:1997,Zhu:2013upa,Zhu:2014aea,Zhu:2016srz,Pan:2023aiv},
\bqn
\lb{standard_eom}
\frac{d^2\nu_k(y)}{dy^2}&=&\left(-1-\frac{m^2_\mathrm{eff}}{k^2}\right)\nu_k(y)\nb\\
&\equiv& \left(\lambda_{\rm U}^2g(y) + q(y)\right) \nu_k (y),
\eqn
where $\lambda_{\rm U}$ is a formal bookkeeping parameter introduced to
organize the UAA hierarchy. It is set to unity only after the asymptotic
expansion has been constructed and should not be interpreted as an
expansion parameter in inverse powers of the comoving wavenumber $k$.

Following the standard Liouville transformation, we introduce a new
independent variable $\xi$ and a new function $U(\xi)$ through
\cite{Olver:1997,Zhu:2013upa}
\bqn
\lb{eq4.5}
\nu_k(y) = \dot{y}^{1/2} U(\xi(y)),\quad \dot{y}^2 g(y) = {\cal F}(\xi),
\eqn
where, in this subsection, an overdot on $y$ denotes differentiation
with respect to $\xi$, namely $\dot y\equiv dy/d\xi$. With this transformation, Eq.~(\ref{standard_eom}) becomes
\bq
\lb{eq3.UAAU}
\frac{d^2U}{d\xi^2}=\left(\lambda_{\rm U}^2\dot y^2g+\psi\right)U,
\eq
with
\bq
\lb{eq4.7}
\psi(y)\equiv \dot y^2q(y)+\dot y^{1/2}\frac{d^2}{d\xi^2}\left(\dot y^{-1/2}\right).
\eq

\begin{figure}[h!]
\includegraphics[width=8cm]{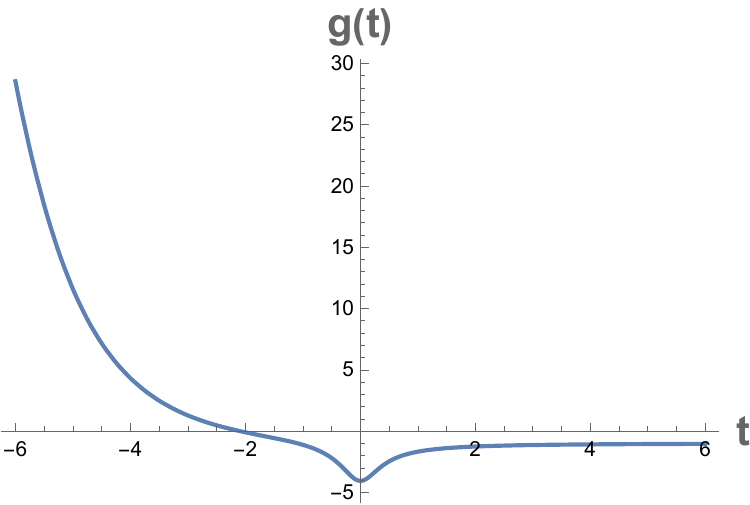}
\caption{The function $g[y(t)]$ defined by Eq.~(\ref{eq3.gdef}) for
$k=0.5\,m_P$. Other values of $k$ exhibit the same qualitative
behavior. In particular, $g[y(t)]$ possesses a single simple zero,
corresponding to the turning point.}
\label{fig3}
\end{figure}

\begin{figure}[t!]
\includegraphics[width=8cm]{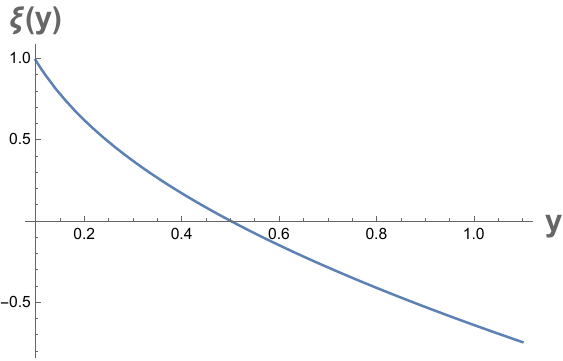}
\caption{The function $\xi(y)$ defined by Eq.~(\ref{eq4.6b}) in the
neighborhood of the turning point for $k=0.5\,m_P$. It decreases
monotonically through $\xi(y_0)=0$ at $y_0\simeq0.5$.} 
\label{fig4}
\end{figure}

In the remote contracting regime, $t\ll t_2$, the asymptotic form
 ~(\ref{eq3.10}) gives
\bqn
\lb{eq3.preDSUAA}
\frac{d^2\nu_k}{dy^2}\simeq\left(-1+\frac{\epsilon_\Lambda}{y^2}\right)\nu_k.
\eqn
To ensure that the UAA error-control function remains finite as
$y\rightarrow0$, we choose \cite{Olver:1997}
\bq
\lb{eq3.qchoice}
q(y)=-\frac{1}{4y^2},
\eq
and retain this choice throughout the pre-inflationary region.
Equation~(\ref{standard_eom}) then gives 
\bqn
\lb{eq3.gdef}
\lambda_{\rm U}^2g(y)&=&-\left(1+\frac{m^2_{\rm eff}}{k^2}+q(y)\right)\nb\\
&=&\frac{1}{4y^2}-1-\frac{m^2_{\rm eff}}{k^2}.
\eqn
Figure~\ref{fig3} shows that $g(y)$ possesses a single simple turning
point $y_0$, defined by $g(y_0)=0$. We therefore choose the Airy
comparison function ${\cal F}(\xi)=\xi$, so that 
\bq
\lb{eq4.9}
\dot y^2g=\xi.
\eq
 With this convention, $\xi$ and $g$ have the same sign, and the
turning point is mapped to $\xi(y_0)=0$, as showmn in Fig. \ref{fig4}.

The corresponding leading-order Airy approximation is  
\bq
\lb{analytical_sol1}
\nu_k^{(1)}(y)=\left(\frac{\xi}{g}\right)^{1/4}
\left\{\tilde a_k\,\mathrm{Ai}(\xi)+\tilde b_k\,\mathrm{Bi}(\xi)\right\}.
\eq
The mapping between $y$ and $\xi$ follows from Eq.~(\ref{eq4.9}) and
satisfies  
\bq
\lb{eq5.6}
\sqrt{|\xi|}\,d\xi=-\sqrt{|g(y)|}\,dy.
\eq
Equivalently,
\bqn
\lb{eq4.6b}
\xi(y)=
\begin{cases}
\displaystyle
\left[
\frac{3}{2}\int_y^{y_0}\sqrt{g(\bar y)}\,d\bar y
\right]^{2/3},
& y\le y_0,\\[3mm]
\displaystyle
-\left[
\frac{3}{2}\int_{y_0}^{y}\sqrt{-g(\bar y)}\,d\bar y
\right]^{2/3},
& y\ge y_0.
\end{cases}
\eqn
This construction ensures that the transformation remains regular at the
simple turning point:
\bqn
\lb{eq4.6aa}
\xi(y_0)=0,\quad 
\left.\frac{\xi}{g}\right|_{y=y_0}=\text{finite and nonzero}.
\eqn

\subsection{Complete Third-Order UAA and Numerical Accuracy}
\label{sec:thirdUAAaccuracy}

For quantitative applications, we retain the higher-order terms of the
one-turning-point UAA developed in Ref.~\cite{Zhu:2014aea}. The general
$2n+1$-order approximation takes the form 
\begin{widetext}
\begin{align}
\nu_k(y)=\left(\frac{\xi}{g}\right)^{1/4}\Bigg\{&\tilde a_k\left[
\mathrm{Ai}(\lambda_{\rm U}^{2/3}\xi)\sum_{s=0}^{n}\frac{A_s(\xi)}{\lambda_{\rm U}^{2s}}
+\frac{\mathrm{Ai}'(\lambda_{\rm U}^{2/3}\xi)}{\lambda_{\rm U}^{4/3}}
\sum_{s=0}^{n-1}\frac{B_s(\xi)}{\lambda_{\rm U}^{2s}}\right]\nonumber\\
+&\tilde b_k\left[
\mathrm{Bi}(\lambda_{\rm U}^{2/3}\xi)\sum_{s=0}^{n}\frac{A_s(\xi)}{\lambda_{\rm U}^{2s}}
+\frac{\mathrm{Bi}'(\lambda_{\rm U}^{2/3}\xi)}{\lambda_{\rm U}^{4/3}}
\sum_{s=0}^{n-1}\frac{B_s(\xi)}{\lambda_{\rm U}^{2s}}\right]\Bigg\},
\label{eq:highUAA}
\end{align}
\end{widetext}
with 
\begin{align}
B_s(\xi)&=\frac{\pm1}{2(\pm\xi)^{1/2}}
\int_0^\xi\frac{\psi(v)A_s(v)-A_s''(v)}{(\pm v)^{1/2}}\,dv,
\nonumber\\
A_{s+1}(\xi)&=-\frac12B_s'(\xi)
+\frac12\int^\xi\psi(v)B_s(v)\,dv , 
\label{eq:UAArecursion}
\end{align}
where $A_0=1$, and the upper and lower signs correspond to
$\xi>0$ and $\xi<0$, respectively.  The complete third-order UAA corresponds to $n=1$. After setting the
formal bookkeeping parameter $\lambda_{\rm U}=1$, it becomes 
\begin{align}
\nu_k^{(3)}=\left(\frac{\xi}{g}\right)^{1/4}\Big\{&\tilde a_k\left[(1+A_1)\mathrm{Ai}(\xi)+B_0\mathrm{Ai}'(\xi)\right]\nonumber\\
+&\tilde b_k\left[(1+A_1)\mathrm{Bi}(\xi)+B_0\mathrm{Bi}'(\xi)\right]\Big\}.
\label{eq:thirdUAA}
\end{align}
Although the coefficient functions $B_0(\xi)$ and $A_1(\xi)$ appear
singular at $\xi=0$ when written in their integral forms, these
singularities are removable. In our numerical implementation, we evaluate
$B_0$ and $A_1$ using local Taylor expansions about the turning point and
join these regular representations smoothly to the integral expressions
away from $\xi=0$.

To assess the accuracy of the UAA, we choose the de Sitter vacuum 
for the mode function given by  Eq.~(\ref{sol3B}), that is, 
$a_k=\sqrt{\pi/(4k)}$ and $b_k=0$. The integration constants of both
the first- and complete third-order UAA solutions are fixed using the
asymptotic pre-de Sitter junction conditions derived in Sec.~III E.
Throughout this procedure we retain the general index
$\mu=(1/4-\epsilon_\Lambda)^{1/2}$ rather than setting
$\mu=1/2$. In particular, the finite part of the Liouville phase
integral and the asymptotic values of $A_1$ and $B_0$ are retained.
Numerically, these asymptotic quantities are evaluated only after they
have converged in the remote contracting branch.

The reference solutions are obtained by direct numerical integration of
Eq.~(\ref{MS_equation}) from $t_i=-20\,t_P$, using the same analytic
pre-de Sitter mode as initial data.

To quantify the comparison, we define the normalized complex-mode RMS
error by
\begin{equation}
{\cal E}_{\nu}\equiv
\frac{\left\langle|\nu_k^{\rm app}-\nu_k^{\rm num}|^2\right\rangle^{1/2}}
{\max|\nu_k^{\rm num}|},
\label{eq:modeerror}
\end{equation}
where $\nu_k^{\rm app}$ denotes the approximate mode function under
consideration, while $\nu_k^{\rm num}$ denotes the reference solution
obtained by direct numerical integration with the same pre-de Sitter
initial data. The brackets $\langle\cdots\rangle$ denote an average over
the numerical evolution interval used in the comparison, and
$\max|\nu_k^{\rm num}|$ is the maximum numerical mode amplitude over
that same interval. This normalization remains well defined even when
the real or imaginary part of the mode crosses zero.

We also define the analogous error in the squared mode amplitude by
\bq
{\cal E}_{|\nu|^2}\equiv
\frac{
\left\langle
\left(|\nu_k^{\rm app}|^2-|\nu_k^{\rm num}|^2\right)^2
\right\rangle^{1/2}
}{
\max |\nu_k^{\rm num}|^2
}.
\eq

\begin{figure}[htp]
{\includegraphics[width=7cm]{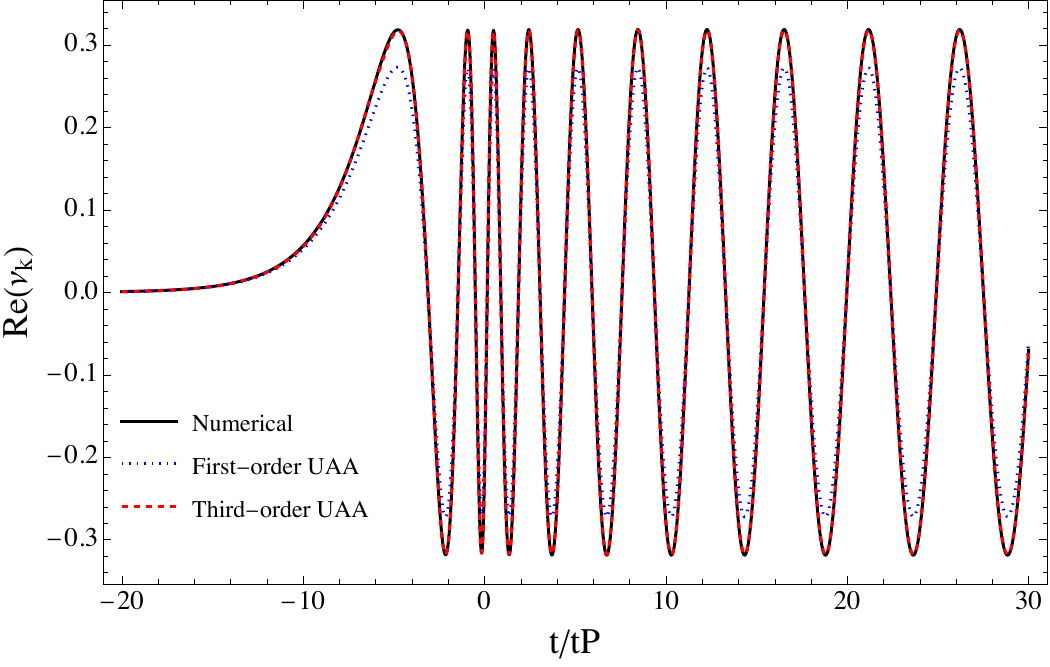}\\
(a)\\
\includegraphics[width=7cm]{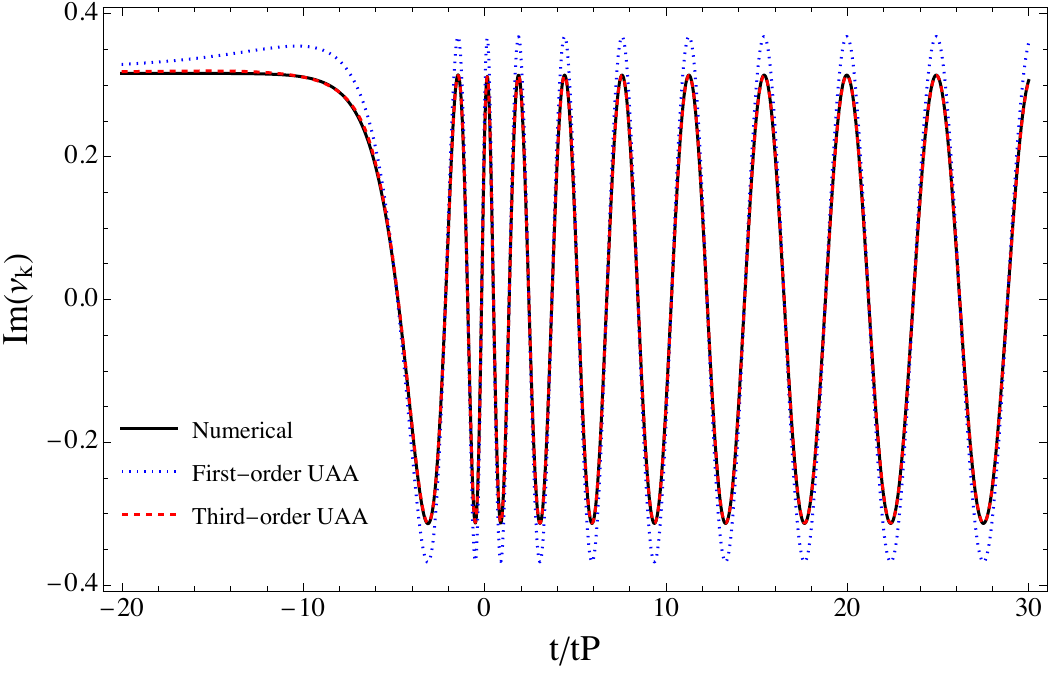}\\
(b)\\}
\caption{Comparison of the numerical solution (solid curve), the first-order UAA (dotted curve), and the complete third-order UAA (dashed curve) for $k=5\,m_P$, with both UAA solutions fixed by the same asymptotic pre-de Sitter junction conditions.  The turning point is $y_0\simeq0.503737$, corresponding to $t_0\simeq-7.51792\,t_P$.  Panels (a) and (b) show the real and imaginary parts, respectively.}
\label{fig5}
\end{figure}

\begin{figure}[htp]
{\includegraphics[width=7cm]{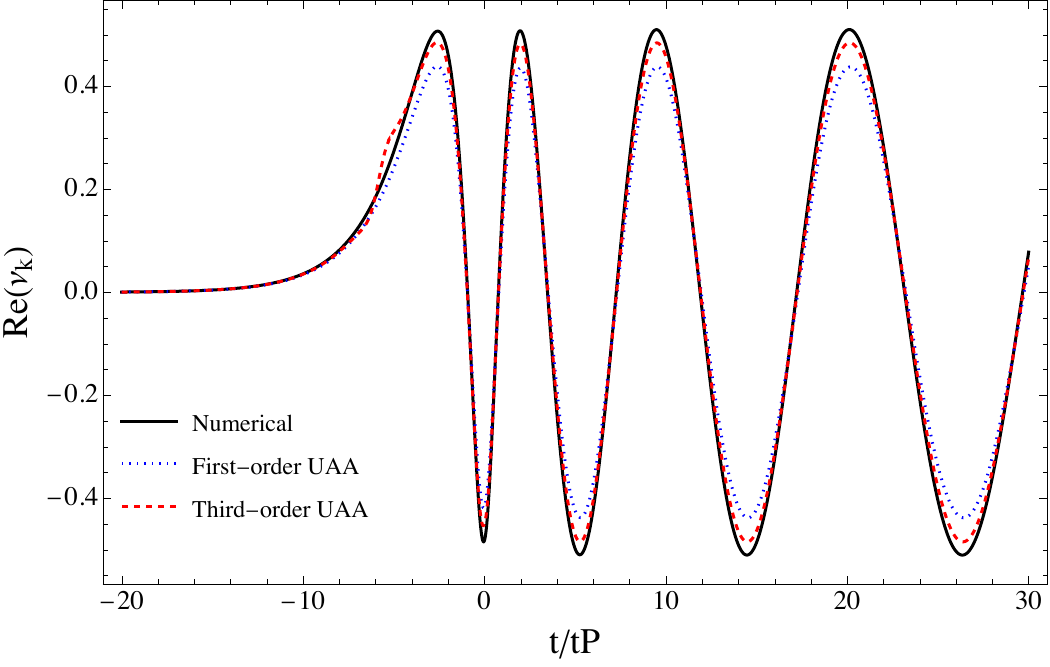}\\
(a)\\
\includegraphics[width=7cm]{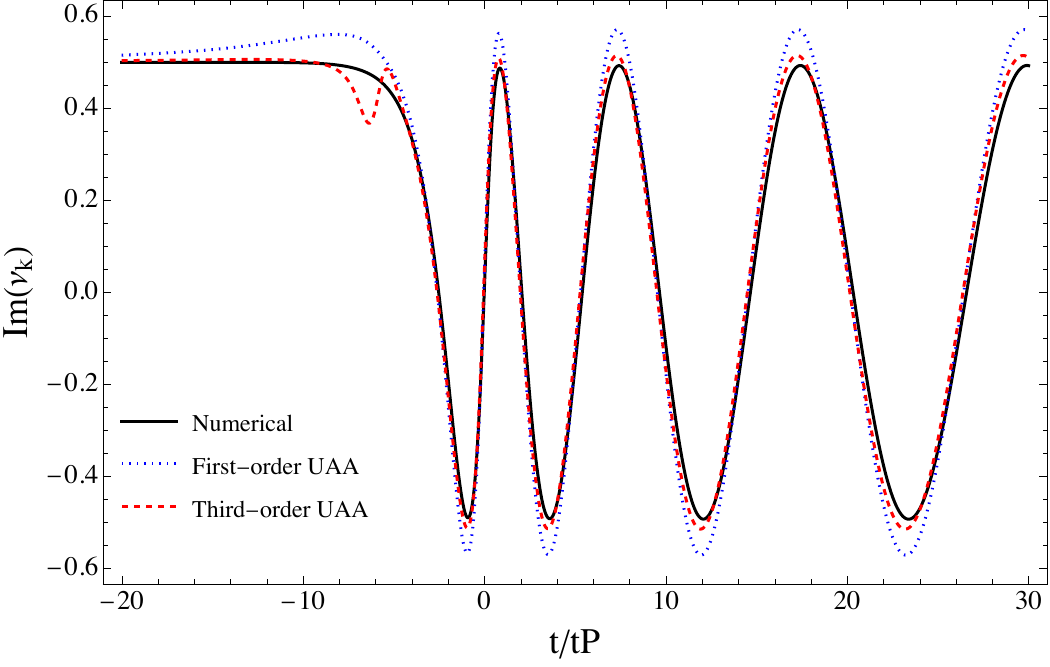}\\
(b)\\}
\caption{Comparison of the numerical solution (solid curve), the first-order UAA (dotted curve), and the complete third-order UAA (dashed curve) for $k=2\,m_P$, with both UAA solutions fixed by the same asymptotic pre-de Sitter junction conditions.  The turning point is $y_0\simeq0.513071$, corresponding to $t_0\simeq-5.27302\,t_P$.  The low-wavenumber mode remains visibly more difficult for the ordinary Airy-UAA hierarchy even after the third-order correction.}
\label{fig6}
\end{figure}

\begin{figure}[htp]
{\includegraphics[width=7cm]{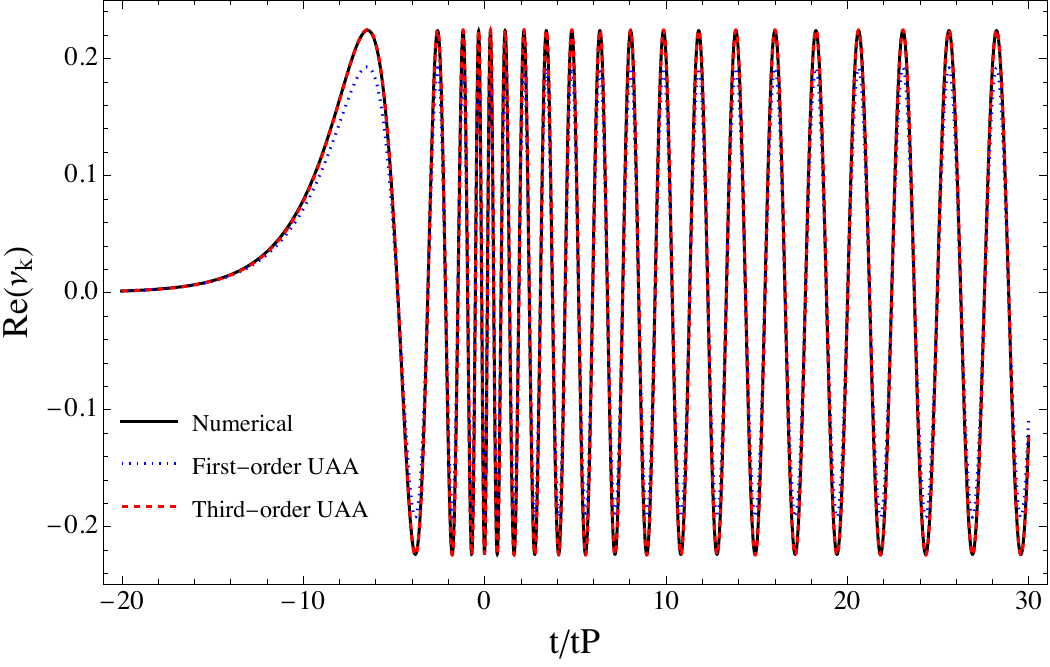}\\
(a)\\
\includegraphics[width=7cm]{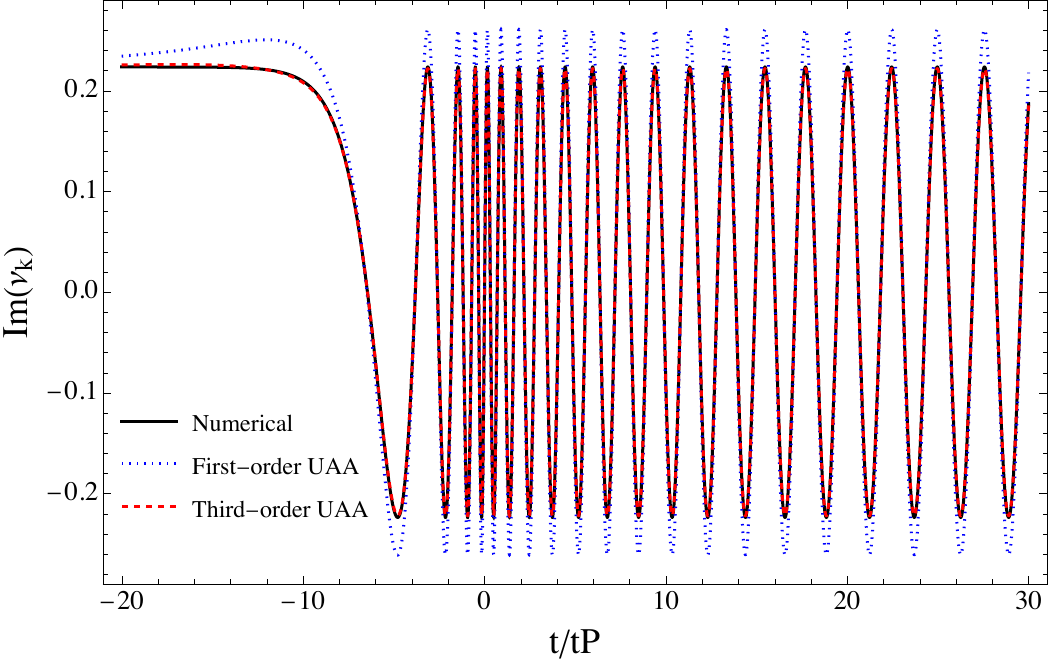}\\
(b)\\}
\caption{Comparison of the numerical solution (solid curve), the first-order UAA (dotted curve), and the complete third-order UAA (dashed curve) for $k=10\,m_P$, with both UAA solutions fixed by the same asymptotic pre-de Sitter junction conditions.  The turning point is $y_0\simeq0.500430$, corresponding to $t_0\simeq-9.19858\,t_P$.}
\label{fig7}
\end{figure}


Figures~\ref{fig5}--\ref{fig7} compare the numerical evolution with
the first- and complete third-order UAA solutions, with all coefficients
fixed by the same asymptotic pre-de Sitter matching prescription.

The leading Airy approximation captures the overall turning-point
behavior but exhibits noticeable amplitude and phase errors. Its
normalized complex-mode RMS errors are approximately $13.5\%$,
$14.4\%$, and $14.4\%$ for $k/m_P=2$, $5$, and $10$,
respectively.
The complete third-order correction substantially improves the
agreement. For the three representative modes we find
\begin{equation}
\begin{array}{c|ccc}
 k/m_P & 2 & 5 & 10\\ \hline
100\,{\cal E}_{\nu}^{(3)} & 7.20 & 0.784 & 0.653\\
100\,{\cal E}_{|\nu|^2}^{(3)} & 9.36 & 0.938 & 1.10
\end{array}
\label{eq:thirdordererrors}
\end{equation}
Thus, the complete third-order UAA reaches sub-percent complex-mode
accuracy for the representative $k=5\,m_P$ and $10\,m_P$ modes,
whereas the $k=2\,m_P$ mode remains appreciably more difficult.
The third-order curves for the two higher-wavenumber examples nearly
overlap the numerical solutions over most of the evolution interval,
while the $k=2\,m_P$ mode retains a visible discrepancy in the strongly
nonadiabatic region.

These results demonstrate both the improvement provided by the
higher-order UAA and the quantitative accuracy of the asymptotic junction
conditions of Sec.~III E for the two higher-wavenumber examples. They
also identify the lower-$k$ sector as the regime in which the ordinary
Airy hierarchy converges more slowly.

The errors quoted above should not be confused with the frequently cited
$\lesssim0.15\%$ accuracy of the third-order UAA reported in
Ref.~\cite{Zhu:2014aea}. That estimate refers to the inflationary power
spectrum in the de Sitter-like benchmark considered there; it is not a
universal pointwise error bound on the complex mode function. The actual
accuracy of the UAA is governed by its error-control functions and
therefore depends on the background and on the turning-point structure.

Thus, the percent-level complex-mode errors found here are fully
compatible with the substantially smaller power-spectrum error reported
for the benchmark considered in Ref.~\cite{Zhu:2014aea}.

\subsection{Fifth-Order Error Control and the Low-$k$ Mode}
\label{sec:fifthUAA}

The next complete approximation in the Olver hierarchy is fifth order,
rather than a stand-alone fourth-order approximation. It contains the
additional coefficient functions $B_1$ and $A_2$ and, after setting
$\lambda_{\rm U}=1$, takes the form 
\begin{align}
\nu_k^{(5)}=\left(\frac{\xi}{g}\right)^{1/4}\Big\{&\tilde a_k\left[(1+A_1+A_2)\mathrm{Ai}+(B_0+B_1)\mathrm{Ai}'\right]\nonumber\\
+&\tilde b_k\left[(1+A_1+A_2)\mathrm{Bi}+(B_0+B_1)\mathrm{Bi}'\right]\Big\}.
\label{eq:fifthUAA}
\end{align}
The remainder associated with this fifth-order approximation is governed
by the next error-control coefficient,
\begin{equation}
B_2(\xi)=\frac{\pm1}{2(\pm\xi)^{1/2}}
\int_0^\xi\frac{\psi(v)A_2(v)-A_2''(v)}{(\pm v)^{1/2}}\,dv,
\label{eq:B2}
\end{equation}
through the variation of $|\xi|^{1/2}B_2(\xi)$ appearing in the rigorous
UAA error bound \cite{Zhu:2014aea}.

 Evaluating $B_2$ at the turning point by means of its regular Taylor
representation gives
\begin{equation}
B_2(0)\simeq
\begin{cases}
18.2, & k=2m_P,\\
1.06\times10^{-2}, & k=5m_P,\\
5.9\times10^{-3}, & k=10m_P.
\end{cases}
\label{eq:B2values}
\end{equation}
The contrast is striking. For the $k=2\,m_P$ mode, the
remainder-control coefficient $B_2(0)$ is more than three orders of
magnitude larger than for the two higher-wavenumber examples. This
provides a clear diagnostic of why simply extending the same Airy
comparison problem to higher formal UAA order is inefficient for this
lower-$k$ mode.

There are two possible ways to improve the approximation in this regime.
A fully analytic approach would be to construct a more adapted Liouville
comparison problem, so that a larger part of the low-$k$ structure is
incorporated into the leading comparison equation and correspondingly
less remains in $\psi$. Such a construction must preserve the pole
structure required for convergence of the UAA error-control functions
and is therefore considerably more involved. A simpler alternative is
to use the more accurate of the available UAA approximations as a seed
and correct only its residual. We adopt this adaptive strategy below.

We emphasize that the UAA is an asymptotic expansion in the formal
bookkeeping parameter $\lambda_{\rm U}$. Since the physical
approximation is obtained after setting $\lambda_{\rm U}=1$, successive
formal orders are not required to improve monotonically. In our direct
implementation of the fifth-order approximation, the resulting global
mode function becomes sensitive to the higher-derivative regularization
and to the matching prescription. We therefore use the fifth-order
analysis primarily as an error-control diagnostic and do not employ
Eq.~(\ref{eq:fifthUAA}) itself as the final precision approximation.

\subsection{Defect-Corrected Approximations}
\label{sec:defect}

The preceding analysis shows that, after the formal bookkeeping parameter
$\lambda_{\rm U}$ has been set to unity, increasing the UAA order does
not necessarily improve the approximation for every wavenumber. We
therefore use the more accurate of the available first- and third-order
UAA solutions as the seed for a defect correction. The two seeds are
compared using the normalized complex-mode RMS error defined in
Eq.~(\ref{eq:modeerror}), and we denote the corresponding errors by
${\cal E}^{(1)}_\nu$ and ${\cal E}^{(3)}_\nu$.

A scan of the low-wavenumber region reveals a crossover at
\begin{equation}
 k_c\simeq0.5457\,m_P,
\label{eq:kc}
\end{equation}
where
\begin{equation}
 {\cal E}_{\nu}^{(1)}(k_c)={\cal E}_{\nu}^{(3)}(k_c)
 \simeq 28.19\%.
\label{eq:kccross}
\end{equation}
For $k<k_c$, the first-order approximation has the smaller normalized
complex-mode RMS error, whereas for $k\geq k_c$ the complete third-order
approximation is preferred. We therefore define the adaptive seed by 
\begin{equation}
\nu_k^{\rm seed}=\begin{cases}
\nu_k^{(1)}, & k<k_c,\\[1mm]
\nu_k^{(3)}, & k\ge k_c.
\end{cases}
\label{eq:adaptiveseed}
\end{equation}
The crossover scale $k_c$ is a practical property of the present
background, matching prescription, comparison interval, and error
measure; it should not be interpreted as a new physical scale.
Notice also that the observationally relevant range in Eq.~(\ref{eq:II-window-comoving})
begins at $k\simeq m_P>k_c$, so that the complete third-order UAA is
the preferred seed throughout that range.  In addition, ${\cal E}_{\nu}^{(3)}$ falls below $15\%$ already at $k\simeq0.853\,m_P$, slightly below the lower edge of the observational window.

The low-$k$ comparison is summarized in Table~\ref{tab:adaptiveerrors}.  The corrected error is defined analogously to Eq.~(\ref{eq:modeerror}) by
\begin{equation}
{\cal E}_{\nu}^{\rm corr}\equiv
\frac{\left\langle\left|\nu_k^{\rm corr}-\nu_k^{\rm num}\right|^2\right\rangle^{1/2}}
{\max|\nu_k^{\rm num}|},
\label{eq:Ecorr}
\end{equation}
where $\nu_k^{\rm corr}$ denotes the total defect-corrected solution
defined below. Thus, ${\cal E}^{\rm corr}_\nu$ measures the numerical
agreement of the complete semi-analytic construction with the direct
solution of the MS equation. It is not a formal UAA truncation error.

\begin{table}[t]
\centering
\begin{tabular}{c|cc|c|c}
\hline\hline
$k/m_P$ & $100{\cal E}_{\nu}^{(1)}$ & $100{\cal E}_{\nu}^{(3)}$
& preferred seed & $100{\cal E}_{\nu}^{\rm corr}$\\
\hline
$0.1$ & $28.52$ & $38.45$ & first & $1.36\times10^{-5}$\\
$0.4$ & $30.00$ & $35.31$ & first & $1.03\times10^{-5}$\\
$0.5$ & $27.75$ & $30.74$ & first & $1.06\times10^{-5}$\\
$k_c/m_P=0.5457$ & $28.19$ & $28.19$ & third & $1.00\times10^{-5}$\\
$0.6$ & $28.51$ & $24.87$ & third & $9.23\times10^{-6}$\\
$0.7$ & $27.09$ & $19.81$ & third & $8.92\times10^{-6}$\\
$0.8$ & $25.13$ & $16.48$ & third & $9.09\times10^{-6}$\\
$0.9$ & $24.29$ & $13.68$ & third & $8.69\times10^{-6}$\\
$1.0$ & $22.26$ & $10.79$ & third & $8.72\times10^{-6}$\\
\hline\hline
\end{tabular}
\caption{Normalized complex-mode RMS errors, in percent, of the first- and complete third-order UAA solutions in the low-wavenumber region.  The preferred seed is the one with the smaller uncorrected error.  The final column gives the error after defect correction.  At the crossover $k_c$ the two seed errors coincide; by convention we use the third-order seed for $k\ge k_c$.}
\label{tab:adaptiveerrors}
\end{table}

Table~\ref{tab:adaptiveerrors} contains the central numerical result underlying the adaptive prescription.  Several features are worth emphasizing.  First, for the very low-wavenumber modes $k=0.1\,m_P$, $0.4\,m_P$, and $0.5\,m_P$, one has ${\cal E}_{\nu}^{(1)}<{\cal E}_{\nu}^{(3)}$.  Thus, after the formal UAA bookkeeping parameter has been set to unity, increasing the UAA order does not necessarily improve the approximation.  Second, the two errors become equal at $k=k_c$, which provides a natural numerical criterion for switching between the two available seeds.  Third, immediately above the crossover the complete third-order approximation becomes preferable: at $k=0.6\,m_P$ the error decreases from $28.51\%$ at first order to $24.87\%$ at third order, and by $k=m_P$ the corresponding reduction is from $22.26\%$ to $10.79\%$.  Since the observationally relevant interval begins at $k\simeq m_P$, the complete third-order UAA is therefore the preferred seed throughout the physical range considered in this work.

The last column of Table~\ref{tab:adaptiveerrors} shows an even more important feature.
Once the residual of the preferred seed is propagated with the exact mode
operator, the corrected error remains at the level of
${\cal O}(10^{-5}\%)$ over the entire interval
$0.1\leq k/m_P\leq1$. Thus, even when the preferred uncorrected seed
has an RMS error of roughly $25$--$30\%$, the defect-corrected solution
agrees with the direct numerical evolution to approximately one part in
$10^7$.  The small variation of ${\cal E}^{\rm corr}_\nu$ among the entries in
the table should not be interpreted as a formal UAA error estimate. At
this level, it is dominated by the numerical accuracy of the residual
propagation, interpolation, and comparison procedure. 
The purpose of the table is instead to demonstrate that the defect correction removes essentially all of the seed error once the same physical de Sitter initial state is imposed on the total solution.

To formulate the defect correction for either seed, define the exact
mode operator
\begin{equation}
Q_k(y)\equiv-1-\frac{m_{\rm eff}^2}{k^2},\quad 
{\cal L}_k\equiv\frac{d^2}{dy^2}-Q_k(y).
\label{eq:defectoperator}
\end{equation}
For a seed of order $s=1$ or $3$, its residual under the exact operator is  
\begin{equation}
R_k^{(s)}(y)\equiv{\cal L}_k\nu_k^{(s)}
=\frac{d^2\nu_k^{(s)}}{dy^2}-Q_k(y)\nu_k^{(s)}.
\label{eq:defectresidual}
\end{equation}
Writing the corrected solution as
\begin{equation}
\nu_k^{\rm corr}=\nu_k^{(s)}+\delta\nu_k^{(s)},
\label{eq:defectsplit}
\end{equation}
and requiring it to satisfy the exact MS equation gives the inhomogeneous
defect equation 
\begin{equation}
\frac{d^2\delta\nu_k^{(s)}}{dy^2}-Q_k(y)\delta\nu_k^{(s)}
=-R_k^{(s)}(y).
\label{eq:defecteq}
\end{equation}
The UAA seed is fixed by the asymptotic junction conditions derived in
Sec.~III E. At the numerical starting point
$y_i=k\eta_i$, corresponding to $t_i=-20\,t_P$, we initialize the
defect by 
\begin{align}
\delta\nu_k^{(s)}(y_i)&=\nu_k^{\rm dS}(y_i)-\nu_k^{(s)}(y_i),\nonumber\\
\delta\nu_{k,y}^{(s)}(y_i)&=\nu_{k,y}^{\rm dS}(y_i)-\nu_{k,y}^{(s)}(y_i),
\label{eq:defectIC}
\end{align}
so that the total corrected solution represents the same incoming de Sitter state as the direct numerical solution.  Equation~(\ref{eq:defecteq}) is then propagated using the exact background coefficient $Q_k(y)$.  

Because ${\cal L}_k$ is the exact mode operator, an exact solution of
the defect equation with the initial data (\ref{eq:defectIC}) would make
$\nu_k^{\rm corr}$ satisfy the original MS equation identically.
Consequently, the residual values
${\cal E}^{\rm corr}_\nu\sim{\cal O}(10^{-5}\%)$ obtained in the
numerical implementation measure the accuracy of the residual
propagation, interpolation, and late-time projection, rather than a new
formal UAA truncation bound. The practical role of the defect construction is to isolate and quantify the correction to the analytic UAA seed while retaining the exact background evolution.

It is also useful to quantify the size of the defect itself relative to
the final corrected mode. We therefore define 
\begin{equation}
{\cal D}_{\nu}^{(s)}\equiv
\frac{\left\langle|\delta\nu_k^{(s)}|^2\right\rangle^{1/2}}
{\left\langle|\nu_k^{\rm corr}|^2\right\rangle^{1/2}}.
\label{eq:defectsize}
\end{equation}
Selected values for the preferred seed are shown in Table ~\ref{tab:defectsize}.
The correction becomes progressively smaller as the seed accuracy improves.  

\begin{table}[t]
\centering
\begin{tabular}{c|c|c}
\hline\hline
$k/m_P$ & preferred seed & $100{\cal D}_{\nu}^{(s)}$\\
\hline
$0.1$ & first & $54.19$\\
$0.4$ & first & $50.10$\\
$0.5$ & first & $41.61$\\
$0.5457$ & third & $40.13$\\
$0.6$ & third & $34.47$\\
$0.8$ & third & $21.42$\\
$1.0$ & third & $12.88$\\
$2.0$ & third & $7.63$\\
\hline\hline
\end{tabular}
\caption{Relative RMS size of the defect correction, defined by Eq.~(\ref{eq:defectsize}), for the preferred UAA seed.  The correction becomes progressively smaller as the seed becomes more accurate.}
\label{tab:defectsize}
\end{table}

Table~\ref{tab:defectsize} complements Table~\ref{tab:adaptiveerrors} by measuring the \emph{size of the correction itself}, rather than only the accuracy of the final corrected solution.  This distinction is important.  At $k=0.1\,m_P$ and $0.4\,m_P$, the preferred first-order seeds require corrections of about $54\%$ and $50\%$, respectively, of the RMS amplitude of the final mode.  Hence, in this deep low-$k$ regime the defect term is not perturbatively small even though the final corrected solution is extremely accurate.  Near the crossover the correction remains substantial, with ${\cal D}_{\nu}^{(3)}\simeq40.1\%$ at $k=k_c$.  It is, however, substantially smaller at larger wavenumbers: about $34.5\%$ at $k=0.6\,m_P$, $12.9\%$ at $k=m_P$, and $7.63\%$ at $k=2\,m_P$.

Taken together, Tables~\ref{tab:adaptiveerrors} and~\ref{tab:defectsize} therefore convey two different aspects of the method.  The first shows that the final defect-corrected solution is uniformly extremely accurate, while the second shows how much work the defect term must do to repair the chosen UAA seed.  The adaptive choice in Eq.~(\ref{eq:adaptiveseed}) selects the lower-error seed on each
side of $k_c$, providing the most accurate available UAA starting point
for the defect propagation.   

For very small $k$ the construction is best regarded as a semi-analytic reconstruction obtained by propagating the residual with the exact operator; for the physically relevant modes the correction becomes progressively smaller and the UAA seed itself provides an increasingly efficient approximation.

\subsubsection{Third-Order UAA Seed: $k\ge k_c$}
\label{sec:defectthird}

\begin{figure}[htp]
{\includegraphics[width=7cm]{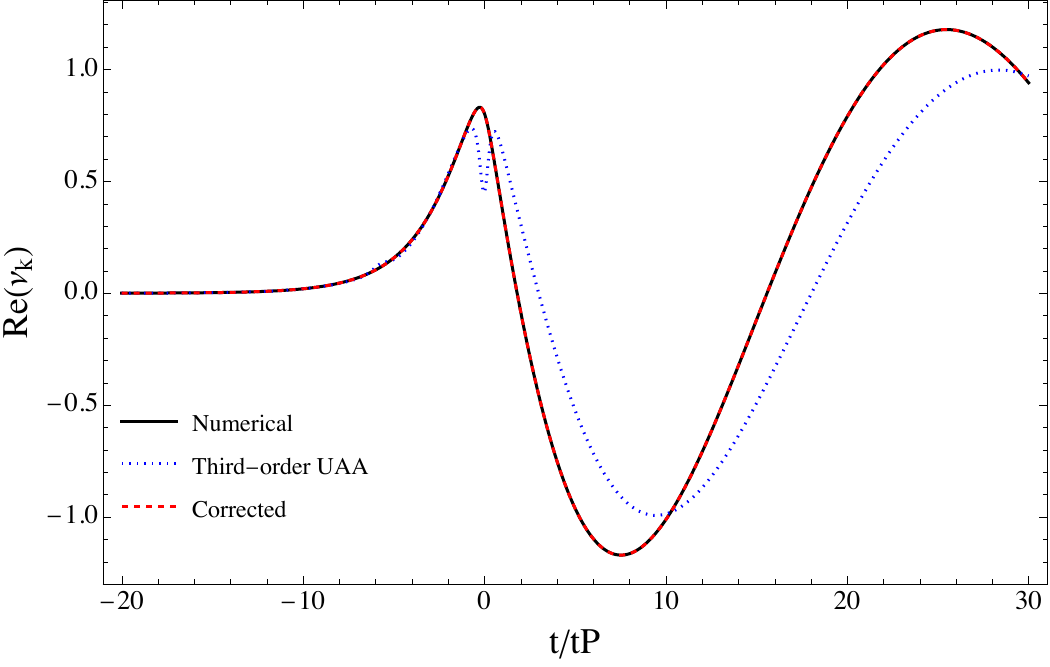}\\
(a)\\
\includegraphics[width=7cm]{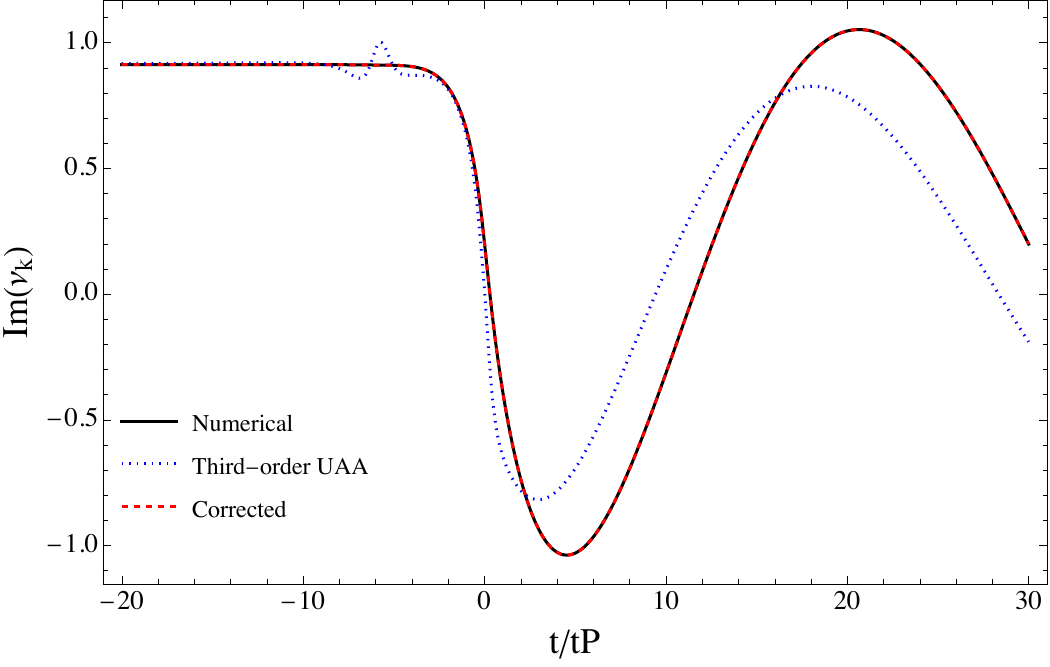}\\
(b)\\}
\caption{Numerical solution, complete third-order UAA seed, and defect-corrected solution for $k=0.6\,m_P>k_c$.  Panels (a) and (b) show the real and imaginary parts, respectively.  The normalized complex-mode RMS error is reduced from $24.87\%$ for the third-order seed to $9.23\times10^{-6}\%$ after defect correction.}
\label{fig:defectk06}
\end{figure}

For $k\ge k_c$, the complete third-order UAA has the smaller RMS error and we set $s=3$ in Eqs.~(\ref{eq:defectresidual})--(\ref{eq:defectIC}).  The representative case $k=0.6\; m_P$, which lies only slightly above the crossover, is deliberately demanding: the uncorrected errors are
\begin{equation}
{\cal E}_{\nu}^{(1)}\simeq28.51\%,\quad 
{\cal E}_{\nu}^{(3)}\simeq24.87\%,
\end{equation}
whereas the defect-corrected result has
\begin{equation}
{\cal E}_{\nu}^{\rm corr}\simeq9.23\times10^{-6}\%.
\end{equation}
Figure~\ref{fig:defectk06} shows that the corrected solution is visually indistinguishable from the direct numerical evolution even though the third-order seed itself still differs substantially.  For the observationally relevant modes $k\ge m_P$, the third-order seed is always the preferred one and its uncorrected accuracy is substantially better than in the near-crossover regime.

\subsubsection{First-Order UAA Seed: $k<k_c$}
\label{sec:defectfirst}

\begin{figure}[htp]
{\includegraphics[width=7cm]{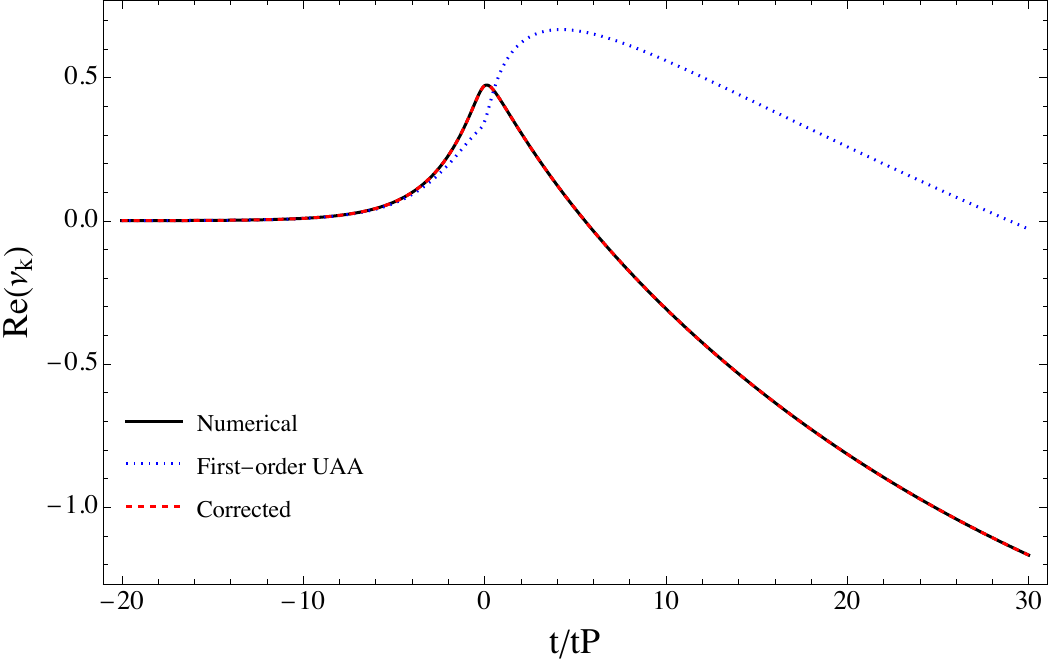}\\
(a)\\
\includegraphics[width=7cm]{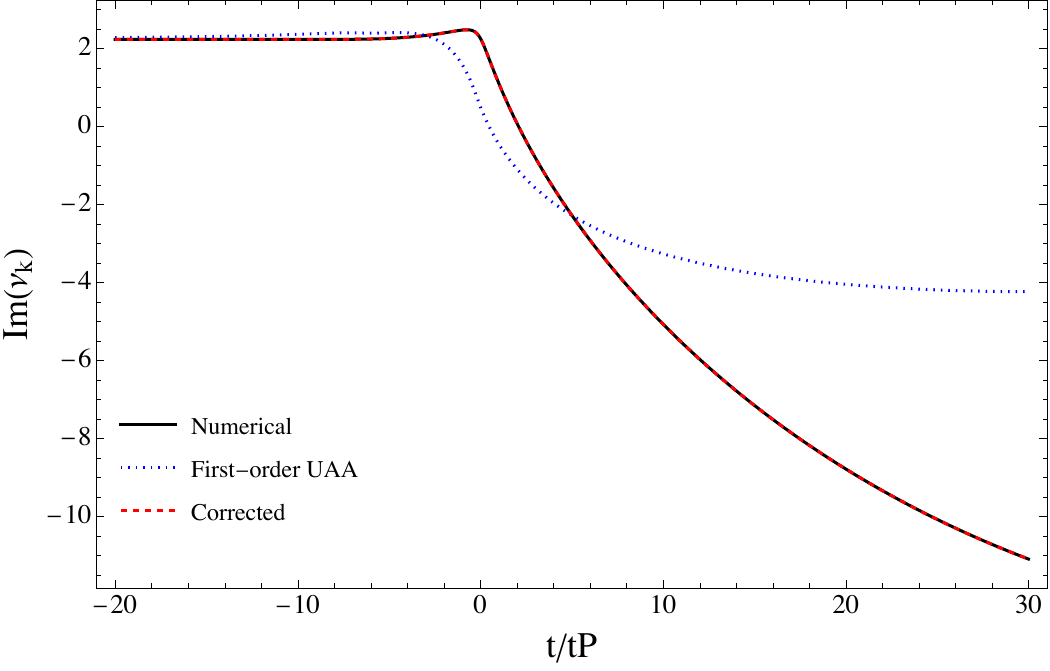}\\
(b)\\}
\caption{Numerical solution, first-order UAA seed, and defect-corrected solution for $k=0.1\,m_P<k_c$.  Panels (a) and (b) show the real and imaginary parts, respectively.  Although the first-order seed has a normalized complex-mode RMS error of $28.52\%$, the defect-corrected solution reduces it to $1.36\times10^{-5}\%$.}
\label{fig:defectk01}
\end{figure}

For $k<k_c$, the leading Airy solution is more accurate than the complete third-order approximation and is therefore used as the seed.  The representative case $k=0.1\; m_P$ provides a stringent stress test well below the observational window.  We find
\begin{equation}
{\cal E}_{\nu}^{(1)}\simeq28.52\%,\quad 
{\cal E}_{\nu}^{(3)}\simeq38.45\%,
\end{equation}
so the first-order seed is clearly preferable.  Nevertheless, after solving the defect equation the error falls to
\begin{equation}
{\cal E}_{\nu}^{\rm corr}\simeq1.36\times10^{-5}\%.
\end{equation}
As shown in Fig.~\ref{fig:defectk01}, the corrected solution again follows the numerical mode essentially exactly.  Here ${\cal D}_{\nu}^{(1)}\simeq54.2\%$, so the defect term is not a small perturbative correction. The significance of this example is instead that the exact-operator
residual propagation remains numerically stable and accurate even when
the bare UAA seed is poor. 

Because Eq.~(\ref{eq:defecteq}) uses the exact mode operator and the total corrected solution is initialized to the same de Sitter state, different seeds converge to the same physical solution up to numerical error. The adaptive choice in Eq.~(\ref{eq:adaptiveseed}) is therefore an efficiency
criterion: it selects the lower-error UAA seed and provides the most
accurate available starting point for the defect propagation.

\subsection{Finite-Wavenumber Matching Across Different Epochs}
\label{sec:IIIE-matching}

The analysis of Sec.~\ref{sec:defect}  shows that the preferred UAA
seed depends on the wavenumber.  For $k\geq k_c$ the complete third-order UAA gives the
smaller normalized mode-function error, whereas for $k<k_c$ the first-order
Airy solution is the better seed.  The matching problem should therefore
follow the same adaptive prescription.  In this subsection we derive the
finite-wavenumber connection formulas separately for these two cases.  In
both branches the general pre-de Sitter index
\begin{equation}
 \mu=\left(\frac14-\epsilon_\Lambda\right)^{1/2}
 \label{eq:IIIE-mu}
\end{equation}
is kept throughout.  The strict ultraviolet limit is different from the
finite-$k$ matching problem considered here and will be discussed separately
in the next section.

For later use, we define the finite parts of the Liouville phase integrals by
\begin{equation}
 {\cal I}_{+}\equiv
 \lim_{y\to0^{+}}
 \left[
 \int_{y}^{y_0}\sqrt{g(u)}\,du+\mu\ln y
 \right],
 \label{eq:IIIE-Iplus}
\end{equation}
and
\begin{equation}
 {\cal I}_{-}\equiv
 \lim_{y\to\infty}
 \left[
 \int_{y_0}^{y}\sqrt{-g(u)}\,du-y
 \right].
 \label{eq:IIIE-Iminus}
\end{equation}
These limits are finite because $g(y)\simeq\mu^2/y^2$ as
$y\to0^+$ and $g(y)\to-1$ as $y\to\infty$. Thus,
\begin{equation}
 \frac23\xi^{3/2}
 =
 -\mu\ln y+{\cal I}_{+} + {\cal{O}}\left(1\right),  
\quad y\to 0^{+},
 \label{eq:IIIE-xiplus}
\end{equation}
while
\begin{equation}
 \frac23(-\xi)^{3/2}
 =
 y+{\cal I}_{-}+ {\cal{O}}\left(1\right),
\quad y\to\infty.
 \label{eq:IIIE-ximinus}
\end{equation}
It is also convenient to introduce
\begin{equation}
 \Phi_{-}\equiv \frac{\pi}{4}-{\cal I}_{-},
\quad
 c_\mu\equiv\cot(\pi\mu).
 \label{eq:IIIE-Phiminus}
\end{equation}
For general noninteger $0<\mu<1$, the small-$y$ expansion of the
pre-de Sitter solution in Eq.~(\ref{sol3B}) is
\begin{align}
 \nu_k^{\rm dS}(y)
 \simeq\;&
 \frac{2^{-\mu}}{\Gamma(\mu+1)}
 \left[
 (a_k+b_k)
 -i c_\mu (a_k-b_k)
 \right] y^{1/2+\mu}
 \nonumber\\
 &+
 i\,\frac{2^\mu\Gamma(\mu)}{\pi}
 (a_k-b_k)y^{1/2-\mu}
 +\cdots .
 \label{eq:IIIE-Hankel-small-y}
\end{align}

\subsubsection{Third-Order UAA Seed: $k\geq k_c$}
\label{sec:IIIE-third}

For $k\geq k_c$, the complete third-order UAA of Eq.~(\ref{eq:thirdUAA}) is used as the
seed.  We define its asymptotic coefficient combinations \footnote{With the normalization adopted for the UAA recursion, these quantities are
related to the asymptotic error-control functions defined in \cite{Zhu:2013upa} by
\begin{align}
 {\cal D}_{+}&=\frac{{\cal H}_{\rm err}(+\infty)}{2},
\quad
 A_1^{+}=\frac12{\cal D}_{+}^{\,2},\nonumber\\
 {\cal D}_{-}&=-\frac{{\cal H}_{\rm err}(-\infty)}{2},
\quad
 A_1^{-}=-\frac12{\cal D}_{-}^{\,2}.
 \label{eq:IIIE-CD-Herr}
\end{align}}
\begin{align}
 {\cal C}_{+}&\equiv 1+A_1^{+}, \quad
 A_1^{+} \equiv\lim_{\xi\to+\infty}A_1(\xi),\nonumber\\
 {\cal D}_{+}&\equiv
 \lim_{\xi\to+\infty}\sqrt{\xi}\,B_0(\xi),
 \nonumber\\
 {\cal C}_{-}&\equiv 1+A_1^{-}, \quad
 A_1^{-}\equiv\lim_{\xi\to-\infty}A_1(\xi),\nonumber\\
 {\cal D}_{-}&\equiv
 \lim_{\xi\to-\infty}\sqrt{-\xi}\,B_0(\xi).
 \label{eq:IIIE-CDpm}
\end{align}

Using the large-positive-$\xi$ asymptotics of the Airy functions in
Eq.~(\ref{eq:thirdUAA}), together with Eq.~(\ref{eq:IIIE-xiplus}), one obtains
\begin{align}
 \nu_k^{(3)}(y)
 &\simeq
 \frac{1}{\sqrt{\pi\mu}}
 \left[
 \frac{\widetilde a_k^{(3)}}{2}
 ({\cal C}_{+}-{\cal D}_{+})
 e^{-{\cal I}_{+}}y^{1/2+\mu}\right.\nonumber\\
 &\quad \left. +
 \widetilde b_k^{(3)}
 ({\cal C}_{+}+{\cal D}_{+})
 e^{{\cal I}_{+}}y^{1/2-\mu}
 \right].
 \label{eq:IIIE-third-small-y}
\end{align}
Matching Eq.~(\ref{eq:IIIE-third-small-y}) to
Eq.~(\ref{eq:IIIE-Hankel-small-y}) yields
\begin{align}
 \widetilde a_k^{(3)}
 &=
 {\cal F}^{(3)}_\mu
 \left[
 (1-i c_\mu)a_k+(1+i c_\mu)b_k
 \right],
 \nonumber\\
 \widetilde b_k^{(3)}
 &=
 i\,{\cal G}^{(3)}_\mu(a_k-b_k),
 \label{eq:IIIE-third-pre-match}
\end{align}
where
\begin{align}
 {\cal F}^{(3)}_\mu
 &\equiv
 \frac{2^{1-\mu}\sqrt{\pi\mu}}{\Gamma(\mu+1)}
 \frac{e^{{\cal I}_{+}}}
 {{\cal C}_{+}-{\cal D}_{+}},
 \nonumber\\
 {\cal G}^{(3)}_\mu
 &\equiv
 2^\mu\Gamma(\mu)\sqrt{\frac{\mu}{\pi}}
 \frac{e^{-{\cal I}_{+}}}
 {{\cal C}_{+}+{\cal D}_{+}} .
 \label{eq:IIIE-FG-third}
\end{align}

We next match the third-order seed to the oscillatory inflationary solution.
Using Eq.~(\ref{eq:IIIE-ximinus}) in the large-negative-$\xi$ Airy
asymptotics gives
\begin{align}
 \nu_k^{(3)}(y)\simeq
 \frac{1}{2\sqrt{\pi}}
 \Big\{&
 \left[
 ({\cal C}_{-}+i{\cal D}_{-})\widetilde a_k^{(3)}
 +({\cal D}_{-}-i{\cal C}_{-})\widetilde b_k^{(3)}
 \right]  \nonumber\\
&\times e^{-iy+i\Phi_{-}}
 \nonumber\\
 &+
 \left[
 ({\cal C}_{-}-i{\cal D}_{-})\widetilde a_k^{(3)}
 +({\cal D}_{-}+i{\cal C}_{-})\widetilde b_k^{(3)}
 \right] \nonumber\\
&\times e^{iy-i\Phi_{-}}
 \Big\}.
 \label{eq:IIIE-third-large-y}
\end{align}
Comparison with Eq.~(\ref{sol3A}) gives the third-order seed coefficients
\begin{align}
 \alpha_k^{(3)}
 &=
 \sqrt{\frac{k}{2\pi}}\,e^{i\Phi_{-}}
 \left[
 ({\cal C}_{-}+i{\cal D}_{-})\widetilde a_k^{(3)}
 +({\cal D}_{-}-i{\cal C}_{-})\widetilde b_k^{(3)}
 \right],
 \nonumber\\
 \beta_k^{(3)}
 &=
 \sqrt{\frac{k}{2\pi}}\,e^{-i\Phi_{-}}
 \left[
 ({\cal C}_{-}-i{\cal D}_{-})\widetilde a_k^{(3)}
 +({\cal D}_{-}+i{\cal C}_{-})\widetilde b_k^{(3)}
 \right].
 \label{eq:IIIE-third-alpha-beta}
\end{align}
Equations~(\ref{eq:IIIE-third-pre-match}) and
(\ref{eq:IIIE-third-alpha-beta}) provide the complete third-order
finite-wavenumber connection from $(a_k,b_k)$ to
$\left(\alpha_k^{(3)},\beta_k^{(3)}\right)$.

For later applications it is useful to write this connection in matrix form,
\begin{equation}
 \begin{pmatrix}
 \alpha_k^{(3)}\\[1mm]
 \beta_k^{(3)}
 \end{pmatrix}
 =
 T_k^{(3)}
 \begin{pmatrix}
 a_k\\[1mm]
 b_k
 \end{pmatrix},
\quad
 T_k^{(3)}=M_-^{(3)}M_+^{(3)},
 \label{eq:IIIE-Tthird}
\end{equation}
with
\begin{equation}
 M_+^{(3)}
 =
 \begin{pmatrix}
 {\cal F}^{(3)}_\mu(1-i c_\mu) &
 {\cal F}^{(3)}_\mu(1+i c_\mu)
 \\
 i{\cal G}^{(3)}_\mu &
 -i{\cal G}^{(3)}_\mu
 \end{pmatrix},
 \label{eq:IIIE-Mplus-third}
\end{equation}
and
\begin{equation}
 M_-^{(3)}
 =
 \sqrt{\frac{k}{2\pi}}
 \begin{pmatrix}
 e^{i\Phi_-}({\cal C}_{-}+i{\cal D}_{-}) &
 e^{i\Phi_-}({\cal D}_{-}-i{\cal C}_{-})
 \\
 e^{-i\Phi_-}({\cal C}_{-}-i{\cal D}_{-}) &
 e^{-i\Phi_-}({\cal D}_{-}+i{\cal C}_{-})
 \end{pmatrix}.
 \label{eq:IIIE-Mminus-third}
\end{equation}
This representation keeps both amplitudes and relative phases and is therefore
particularly convenient for future calculations of the primordial power
spectrum and non-Gaussian correlation functions.

The corresponding defect-corrected inflationary coefficients are obtained by
writing
\begin{equation}
 \nu_k^{\rm corr}
 =
 \nu_k^{(3)}+\delta\nu_k^{(3)}.
 \label{eq:IIIE-third-corr-split}
\end{equation}
In the oscillatory region, where
\begin{equation}
 \delta\nu_k^{(3)}
 =
 \frac{1}{\sqrt{2k}}
 \left(
 \delta\alpha_k^{(3)}e^{-iy}
 +
 \delta\beta_k^{(3)}e^{iy}
 \right),
 \label{eq:IIIE-third-delta-mode}
\end{equation}
the corrections to the Bogoliubov coefficients are 
\begin{align}
 \delta\alpha_k^{(3)}
 &=
 \sqrt{\frac{k}{2}}\,e^{iy}
 \left(
 \delta\nu_k^{(3)}
 +i\,\partial_y\delta\nu_k^{(3)}
 \right),
 \nonumber\\
 \delta\beta_k^{(3)}
 &=
 \sqrt{\frac{k}{2}}\,e^{-iy}
 \left(
 \delta\nu_k^{(3)}
 -i\,\partial_y\delta\nu_k^{(3)}
 \right).
 \label{eq:IIIE-third-delta-alpha-beta}
\end{align}
Thus
\begin{equation}
 \alpha_k^{\rm corr}
 =
 \alpha_k^{(3)}+\delta\alpha_k^{(3)},
\quad
 \beta_k^{\rm corr}
 =
 \beta_k^{(3)}+\delta\beta_k^{(3)}.
 \label{eq:IIIE-third-corrected-ab}
\end{equation}
Equivalently, the defect contribution can be organized as a correction to the seed transfer matrix,
\begin{equation}
 T_k^{\rm corr}=T_k^{(3)}+\Delta T_k^{(3)},
\; (k\geq k_c).
 \label{eq:IIIE-third-Tcorr}
\end{equation}

\subsubsection{First-Order UAA Seed: $k<k_c$}
\label{sec:IIIE-first}

For $k<k_c$, Sec.~\ref{sec:defect} shows that the leading Airy approximation is the
more accurate seed.  The matching must therefore be carried out directly
with
\begin{equation}
 \nu_k^{(1)}(y)
 =
 \left(\frac{\xi}{g}\right)^{1/4}
 \left[
 \widetilde a_k^{(1)}\,{\rm Ai}(\xi)
 +
 \widetilde b_k^{(1)}\,{\rm Bi}(\xi)
 \right].
 \label{eq:IIIE-first-seed}
\end{equation}
For $\xi\to+\infty$, Eqs.~(\ref{eq:IIIE-Iplus}) and
(\ref{eq:IIIE-xiplus}) give
\begin{equation}
 \nu_k^{(1)}(y)
 \simeq
 \frac{1}{\sqrt{\pi\mu}}
 \left[
 \frac{\widetilde a_k^{(1)}}{2}
 e^{-{\cal I}_{+}}y^{1/2+\mu}
 +
 \widetilde b_k^{(1)}
 e^{{\cal I}_{+}}y^{1/2-\mu}
 \right].
 \label{eq:IIIE-first-small-y}
\end{equation}
Matching to Eq.~(\ref{eq:IIIE-Hankel-small-y}) gives
\begin{align}
 \widetilde a_k^{(1)}
 &=
 {\cal F}^{(1)}_\mu
 \left[
 (1-i c_\mu)a_k+(1+i c_\mu)b_k
 \right],
 \nonumber\\
 \widetilde b_k^{(1)}
 &=
 i\,{\cal G}^{(1)}_\mu(a_k-b_k),
 \label{eq:IIIE-first-pre-match}
\end{align}
where
\begin{align}
 {\cal F}^{(1)}_\mu
 &\equiv
 \frac{2^{1-\mu}\sqrt{\pi\mu}}{\Gamma(\mu+1)}
 e^{{\cal I}_{+}},
 \nonumber\\
 {\cal G}^{(1)}_\mu
 &\equiv
 2^\mu\Gamma(\mu)\sqrt{\frac{\mu}{\pi}}
 e^{-{\cal I}_{+}}.
 \label{eq:IIIE-FG-first}
\end{align}
These expressions retain the exact pre-de Sitter index and the finite
background-dependent phase integral even though the Airy seed itself is only
first order.

For $\xi\to-\infty$, the leading Airy asymptotics give
\begin{align}
 \nu_k^{(1)}(y)
 &\simeq
 \frac{1}{2\sqrt{\pi}}
 \left[
 \left(
 \widetilde a_k^{(1)}-i\widetilde b_k^{(1)}
 \right)e^{-iy+i\Phi_-}\right.\nonumber\\
 &\quad \left.
 +
 \left(
 \widetilde a_k^{(1)}+i\widetilde b_k^{(1)}
 \right)e^{iy-i\Phi_-}
 \right].
 \label{eq:IIIE-first-large-y}
\end{align}
Comparing with Eq.~(\ref{sol3A}), we find
\begin{align}
 \alpha_k^{(1)}
 &=
 \sqrt{\frac{k}{2\pi}}\,
 e^{i\Phi_-}
 \left(
 \widetilde a_k^{(1)}
 -i\widetilde b_k^{(1)}
 \right),
 \nonumber\\
 \beta_k^{(1)}
 &=
 \sqrt{\frac{k}{2\pi}}\,
 e^{-i\Phi_-}
 \left(
 \widetilde a_k^{(1)}
 +i\widetilde b_k^{(1)}
 \right).
 \label{eq:IIIE-first-alpha-beta}
\end{align}
The corresponding first-order transfer matrix is
\begin{equation}
 \begin{pmatrix}
 \alpha_k^{(1)}\\[1mm]
 \beta_k^{(1)}
 \end{pmatrix}
 =
 T_k^{(1)}
 \begin{pmatrix}
 a_k\\[1mm]
 b_k
 \end{pmatrix},
\quad
 T_k^{(1)}=M_-^{(1)}M_+^{(1)},
 \label{eq:IIIE-Tfirst}
\end{equation}
where
\begin{equation}
 M_+^{(1)}
 =
 \begin{pmatrix}
 {\cal F}^{(1)}_\mu(1-i c_\mu) &
 {\cal F}^{(1)}_\mu(1+i c_\mu)
 \\
 i{\cal G}^{(1)}_\mu &
 -i{\cal G}^{(1)}_\mu
 \end{pmatrix},
 \label{eq:IIIE-Mplus-first}
\end{equation}
and
\begin{equation}
 M_-^{(1)}
 =
 \sqrt{\frac{k}{2\pi}}
 \begin{pmatrix}
 e^{i\Phi_-} & -i e^{i\Phi_-}
 \\
 e^{-i\Phi_-} & i e^{-i\Phi_-}
 \end{pmatrix}.
 \label{eq:IIIE-Mminus-first}
\end{equation}

As a useful algebraic check, if we set
\begin{equation}
 \mu=\frac12,
\quad
 {\cal I}_{+}={\cal I}_{-}=0,
 \label{eq:IIIE-leading-limit}
\end{equation}
we have
\begin{equation}
 \widetilde a_k^{(1)}=2(a_k+b_k),
\quad
 \widetilde b_k^{(1)}=i(a_k-b_k),
 \label{eq:IIIE-first-leading-abtilde}
\end{equation}
and Eq.~(\ref{eq:IIIE-first-alpha-beta}) reduces to
\begin{align}
 \alpha_k^{(1)}
 &=
 \sqrt{\frac{k}{2\pi}}\,
 e^{i\pi/4}(3a_k+b_k),
 \nonumber\\
 \beta_k^{(1)}
 &=
 \sqrt{\frac{k}{2\pi}}\,
 e^{-i\pi/4}(a_k+3b_k),
 \label{eq:IIIE-first-leading-alpha-beta}
\end{align}
which reproduces the familiar leading Airy connection formulas.

The defect-corrected completion proceeds in exact parallel with the
third-order branch.  Writing
\begin{equation}
 \nu_k^{\rm corr}
 =
 \nu_k^{(1)}+\delta\nu_k^{(1)},
 \label{eq:IIIE-first-corr-split}
\end{equation}
we obtain in the oscillatory region
\begin{align}
 \delta\alpha_k^{(1)}
 &=
 \sqrt{\frac{k}{2}}\,e^{iy}
 \left(
 \delta\nu_k^{(1)}
 +i\,\partial_y\delta\nu_k^{(1)}
 \right),
 \nonumber\\
 \delta\beta_k^{(1)}
 &=
 \sqrt{\frac{k}{2}}\,e^{-iy}
 \left(
 \delta\nu_k^{(1)}
 -i\,\partial_y\delta\nu_k^{(1)}
 \right),
 \label{eq:IIIE-first-delta-alpha-beta}
\end{align}
and hence
\begin{equation}
 \alpha_k^{\rm corr}
 =
 \alpha_k^{(1)}+\delta\alpha_k^{(1)},
\quad
 \beta_k^{\rm corr}
 =
 \beta_k^{(1)}+\delta\beta_k^{(1)},
 \label{eq:IIIE-first-corrected-ab}
\end{equation}
which can be also cast in the form 
\begin{equation}
 T_k^{\rm corr}=T_k^{(1)}+\Delta T_k^{(1)},
\; (k<k_c).
 \label{eq:IIIE-first-Tcorr}
\end{equation}

\subsubsection{3. Calculations of $\Delta T_k^{(s)}$} 
\label{sec:DeltaTks}

Combining the two branches, the adaptive finite-wavenumber connection may be
summarized as
\begin{equation}
 T_k
 =
 \begin{cases}
 T_k^{(1)}, & k<k_c,\\[1mm]
 T_k^{(3)}, & k\geq k_c,
 \end{cases}
\;\;
 \Delta T_k
 =
 \begin{cases}
 \Delta T_k^{(1)}, & k<k_c,\\[1mm]
 \Delta T_k^{(3)}, & k\geq k_c,
 \end{cases}
 \label{eq:IIIE-adaptive-transfer-a}
\end{equation}
so that
\begin{equation}
 T_k^{\rm corr}=T_k +\Delta T_k.
 \label{eq:IIIE-adaptive-transfer}
\end{equation}

To make the correction matrix $\Delta T_k^{(s)}$ directly usable, it is convenient to construct it from the two independent incoming basis states.  Let
\begin{equation}
 h_{a,k}(y)\equiv y^{1/2}H_{\mu}^{(2)}(y),\quad
 h_{b,k}(y)\equiv y^{1/2}H_{\mu}^{(1)}(y),
 \label{eq:IIIE-Hankel-basis}
\end{equation}
so that the pre-de Sitter solution is $\nu_k^{\rm dS}=a_k h_{a,k}+b_k h_{b,k}$.  For a given seed order $s=1$ or $3$, define the corresponding UAA basis functions $\phi_{a,k}^{(s)}$ and $\phi_{b,k}^{(s)}$ by choosing $(a_k,b_k)=(1,0)$ and $(0,1)$, respectively.  Then
\begin{equation}
 \nu_k^{(s)}(y)=a_k\phi_{a,k}^{(s)}(y)+b_k\phi_{b,k}^{(s)}(y).
 \label{eq:IIIE-seed-basis-decomp}
\end{equation}
Linearity of the defect equation implies the analogous decomposition
\begin{equation}
 \delta\nu_k^{(s)}(y)=a_k\delta\phi_{a,k}^{(s)}(y)+b_k\delta\phi_{b,k}^{(s)}(y).
 \label{eq:IIIE-defect-basis-decomp}
\end{equation}
For $j=a,b$, the basis correction satisfies
\bqn
 && \frac{d^2\delta\phi_{j,k}^{(s)}}{dy^2}-Q_k(y)\delta\phi_{j,k}^{(s)}
 =-r_{j,k}^{(s)}(y),\nb\\
&& r_{j,k}^{(s)}\equiv {\cal L}_k\phi_{j,k}^{(s)},
 \label{eq:IIIE-basis-defect-eq}
\eqn 
with the initial data
\begin{align}
 \delta\phi_{j,k}^{(s)}(y_i)&=h_{j,k}(y_i)-\phi_{j,k}^{(s)}(y_i),\nonumber\\
 \partial_y\delta\phi_{j,k}^{(s)}(y_i)&=\partial_y h_{j,k}(y_i)-\partial_y\phi_{j,k}^{(s)}(y_i).
 \label{eq:IIIE-basis-defect-IC}
\end{align}
Choose $y_f$ in the late oscillatory region where Eq.~(\ref{sol3A}) applies.  The four entries of the correction matrix are then obtained directly from
\begin{align}
 \delta\alpha_{k,j}^{(s)}&=
 \sqrt{\frac{k}{2}}e^{iy_f}
 \left[\delta\phi_{j,k}^{(s)}(y_f)+i\,\partial_y\delta\phi_{j,k}^{(s)}(y_f)\right],\nonumber\\
 \delta\beta_{k,j}^{(s)}&=
 \sqrt{\frac{k}{2}}e^{-iy_f}
 \left[\delta\phi_{j,k}^{(s)}(y_f)-i\,\partial_y\delta\phi_{j,k}^{(s)}(y_f)\right],
 \label{eq:IIIE-delta-columns}
\end{align}
where $j=a, b$. 
Thus, the matrix that appears in Eq.~(\ref{eq:IIIE-adaptive-transfer}) is 
\begin{equation}
 \Delta T_k^{(s)}=
 \begin{pmatrix}
 \delta\alpha_{k,a}^{(s)} & \delta\alpha_{k,b}^{(s)}\\[1mm]
 \delta\beta_{k,a}^{(s)} & \delta\beta_{k,b}^{(s)}
 \end{pmatrix},
\quad s=1,3.
 \label{eq:IIIE-explicit-DeltaT}
\end{equation}
Equations~(\ref{eq:IIIE-basis-defect-eq})--(\ref{eq:IIIE-explicit-DeltaT}) give a direct computational prescription for the defect-corrected transfer matrix: solve two linear defect problems, one for each independent incoming Hankel basis state, and use their late-time values to construct the two columns of $\Delta T_k^{(s)}$.  A compact propagator representation of the same result, together with a Wronskian consistency check, is given in Appendix~\ref{app:defect-propagator}.

The crossover therefore determines which UAA approximation provides the
more efficient analytic seed, while $\Delta T_k^{(s)}$ supplies the
defect correction to the finite-wavenumber connection.   Since the exact defect equation is linear and the total corrected
solution is initialized to the same incoming pre-de Sitter state, the final
corrected connection is independent of the choice of seed up to the numerical
accuracy of the residual propagation.  In particular, the two branches form
one continuous physical construction even though different UAA orders are
used on opposite sides of $k_c$.

\section{Initial Conditions Imposed in the Remote Contracting Phase, Ultraviolet Regularity and Particle Creation}
\lb{SecIV}

We now impose the initial conditions in the remote contracting branch.
Because the background approaches an effective de Sitter spacetime and the
mode function is given by Eq.~(\ref{sol3B}), a natural choice is the de
Sitter vacuum \cite{Pan:2025ppa},
\footnote{Note the difference between the constants $(a_k,b_k)$ used
here and those in Ref.~\cite{Pan:2025ppa}. They are related by
$(a_k,b_k)=\sqrt{\pi/2}\,(\check a_k,\check b_k)$.}
\begin{equation}
\lb{eq4.1}
a_k=\sqrt{\frac{\pi}{4k}},\quad b_k=0.
\end{equation}
Section~\ref{sec:IIIE-matching} provides a finite-wavenumber connection
through the pre-inflationary region.  For the ultraviolet properties of the
state, however, a different limiting procedure is required.  We first explain
why the finite-$k$ matching cannot by itself determine the ultraviolet
Bogoliubov coefficient, and then carry out the matching directly in a
fourth-order adiabatic basis.  This also provides the appropriate definition
of particle creation in the ultraviolet sector.

\subsection{Why the finite-wavenumber matching does not determine the ultraviolet sector}
\label{sec:IV-finite-vs-UV}

The connection formulas derived in Sec.~\ref{sec:IIIE-matching}  rely, on the
pre-de Sitter side, on matching the pre-de Sitter solution to
the UAA solution in the small-$y$ overlap region, 
\begin{equation}
y=k\eta\ll1.
\label{eq:IV-small-y}
\end{equation}
This is the appropriate limit for following a fixed finite-wavenumber mode
from the remote contracting branch into the pre-inflationary region.  However, the
strict ultraviolet limit relevant to adiabatic regularization is different.
At any fixed nonzero conformal time $\eta$, we have
\begin{equation}
k\rightarrow\infty
\quad \Longrightarrow\quad
y=k\eta\rightarrow\infty.
\label{eq:IV-UV-limit}
\end{equation}
Thus the small-$y$ matching and the fixed-time ultraviolet limit are
different asymptotic limits and cannot be interchanged.

This distinction is especially important for particle creation.  A
finite-order UAA approximation may provide an accurate description of the
full mode function for finite $k$, while still being insufficient to resolve
the much smaller negative-frequency component in the ultraviolet.
Consequently, ordinary percentage accuracy of $\nu_k$ does not by itself
establish the correct large-$k$ behavior of the Bogoliubov coefficient
$\beta_k$.  In particular, the $k$-independent negative-frequency coefficient
obtained by extrapolating the leading Airy connection to $k\rightarrow
\infty$ is not the ultraviolet limit of the exact state; it is an artifact of
applying the finite-$k$ connection formula outside the asymptotic regime in
which it was derived.

The ultraviolet problem must therefore be treated separately.  At fixed
time one must use the large-$k$, equivalently large-$y$, expansion of the
mode functions and match it consistently to the adiabatic expansion through
the order required for renormalization.  We now carry out this matching for
the remote de Sitter state.

\subsection{Fourth-order adiabatic matching of the remote de Sitter state}
\label{sec:IV-dS-adiabatic}

At a fixed time in the remote contracting branch, ultraviolet modes satisfy
$|k\eta|\gg1$.  The large-argument expansion of the Hankel function is
\begin{widetext}
\begin{align}
H_\mu^{(2)}(y)\simeq&
\sqrt{\frac{2}{\pi y}}\,
e^{-i(y-\pi\mu/2-\pi/4)}
\Bigg[
1-i\frac{4\mu^2-1}{8y}
-\frac{(4\mu^2-1)(4\mu^2-9)}{2!(8y)^2}
\nonumber\\
&\hspace{1.5cm}
+i\frac{(4\mu^2-1)(4\mu^2-9)(4\mu^2-25)}{3!(8y)^3}
+\frac{(4\mu^2-1)(4\mu^2-9)(4\mu^2-25)(4\mu^2-49)}
{4!(8y)^4}
+{\cal O}(y^{-5})
\Bigg].
\label{eq:IV-Hankel-large-y}
\end{align}
For $\mu^2=1/4-\epsilon_\Lambda$ with the initial condition of Eq.~(\ref{eq4.1}), and after absorbing the irrelevant constant phase into the definition of the positive-frequency basis,
 we find 
\bq
\nu_k^{\rm dS}(\eta)
\simeq \frac{e^{-ik\eta}}{\sqrt{2k}}
\Bigg[
1+i\frac{\epsilon_\Lambda}{2k\eta}
-\frac{\epsilon_\Lambda(\epsilon_\Lambda+2)}
{8k^2\eta^2}
-i\frac{\epsilon_\Lambda(\epsilon_\Lambda+2)(\epsilon_\Lambda+6)}
{48k^3\eta^3}
+\frac{\epsilon_\Lambda(\epsilon_\Lambda+2)(\epsilon_\Lambda+6)
(\epsilon_\Lambda+12)}
{384k^4\eta^4}
+{\cal O}(k^{-5})
\Bigg].
\label{eq:IV-dS-large-k}
\eq
\end{widetext}

The same expansion is obtained from the fourth-order adiabatic solution.
Writing
\begin{equation}
s(\eta)\equiv m_{\rm eff}^2(\eta),
\end{equation}
the fourth-order adiabatic frequency has the large-$k$ form
\begin{equation}
W_k^{(4)}
=
k+\frac{s}{2k}
-\frac{s^2+s''}{8k^3}
+{\cal O}(k^{-5}),
\label{eq:IV-W4}
\end{equation}
and the corresponding normalized positive-frequency mode is
\begin{equation}
u_k^{(4)}(\eta)
=
\frac{1}{\sqrt{2W_k^{(4)}(\eta)}}
\exp\left[
-i\int^\eta W_k^{(4)}(\widetilde\eta)\,d\widetilde\eta
\right].
\label{eq:IV-u4}
\end{equation}
In the asymptotic de Sitter region,
\begin{equation}
s(\eta)=\frac{\epsilon_\Lambda}{\eta^2},
\quad
s''(\eta)=\frac{6\epsilon_\Lambda}{\eta^4}.
\end{equation}
Expanding Eq.~(\ref{eq:IV-u4}) reproduces
Eq.~(\ref{eq:IV-dS-large-k}) term by term through fourth adiabatic
order.  Hence
\begin{equation}
\nu_k^{\rm dS}(\eta)
=
u_k^{(4)}(\eta)
+{\cal O}(k^{-5}),
\label{eq:IV-dS-WKB-match}
\end{equation}
in the sense of the large-$k$ asymptotic series multiplying the common
factor $e^{-ik\eta}/\sqrt{2k}$.  Differentiating the two asymptotic
expansions shows that their first derivatives also agree through the
corresponding fourth adiabatic order.  Thus, both the mode and the canonical
momentum have the ultraviolet behavior required for fourth-order adiabatic
matching.

At any fixed matching time, the ultraviolet Bogoliubov coefficients
are defined by decomposing the mode and its first derivative with
respect to the fourth-order adiabatic basis, 
\begin{equation}
\nu_k(\eta)
=
\alpha_k^{(4)}(\eta)\,u_k^{(4)}(\eta)
+
\beta_k^{(4)}(\eta)\,u_k^{(4)*}(\eta).
\label{eq:IV-adiabatic-decomp}
\end{equation}
Since $u_k^{(4)}$ is normalized to the same Wronskian as $\nu_k$, the
coefficients at any matching time are obtained directly from the mode and its
first derivative:
\begin{align}
\alpha_k^{(4)}
&=
i\left[
u_k^{(4)*}\nu_k'
-u_k^{(4)*\prime}\nu_k
\right],
\nonumber\\
\beta_k^{(4)}
&=
-i\left[
u_k^{(4)}\nu_k'
-u_k^{(4)\prime}\nu_k
\right].
\label{eq:IV-Bogoliubov-projection}
\end{align}
The Wronskian condition immediately gives
\begin{equation}
\left|\alpha_k^{(4)}\right|^2-\left|\beta_k^{(4)}\right|^2=1.
\label{eq:IV-Bogoliubov-Wronskian}
\end{equation}

Applying Eq.~(\ref{eq:IV-Bogoliubov-projection}) to the remote de Sitter
mode, and using the agreement of both the mode and its first derivative with
the fourth-order adiabatic expansion, one obtains, up to the irrelevant phase
convention,
\begin{equation}
\alpha_{k,i}^{(4)}=1+{\cal O}(k^{-5}),
\quad
\beta_{k,i}^{(4)}={\cal O}(k^{-5}),
\quad k\rightarrow\infty.
\label{eq:IV-initial-alpha-beta}
\end{equation}
Thus, the de Sitter state contains no ultraviolet negative-frequency
component through fourth adiabatic order with respect to this adiabatic
basis.  Equation~(\ref{eq:IV-initial-alpha-beta}) is the appropriate
ultraviolet matching condition and replaces any extrapolation of the
finite-$k$ Airy connection to the strict ultraviolet limit.

\subsection{Exact evolution through the bounce and particle creation}
\label{sec:IV-particle-creation}

Having fixed the initial state in the remote contracting branch, we evolve
the exact Mukhanov--Sasaki equation, Eq.~(\ref{MS_equation}), through the
pre-inflationary region.  Particle creation is then determined by projecting
the evolved exact mode onto a positive/negative-frequency adiabatic basis at
a time $\eta_f$ at which the mode is again in an adiabatic regime.  No Airy
matching coefficient is required in this step.

Specifically, let $u_{k,f}^{(4)}$ denote the fourth-order adiabatic mode
constructed from the exact background at $\eta_f$.  The late-time
Bogoliubov coefficients are
\begin{align}
\alpha_{k,f}^{(4)}
&=
i\left[
u_{k,f}^{(4)*}\nu_k'(\eta_f)
-u_{k,f}^{(4)*\prime}\nu_k(\eta_f)
\right],
\nonumber\\
\beta_{k,f}^{(4)}
&=
-i\left[
u_{k,f}^{(4)}\nu_k'(\eta_f)
-u_{k,f}^{(4)\prime}\nu_k(\eta_f)
\right].
\label{eq:IV-final-alpha-beta}
\end{align}
The particle number associated with this late-time adiabatic basis is
therefore
\begin{equation}
N_k=\left|\beta_{k,f}^{(4)}\right|^2.
\label{eq:IV-particle-number}
\end{equation}
Because the particle interpretation is meaningful only in an adiabatic
regime, Eq.~(\ref{eq:IV-particle-number}) is to be evaluated after the
strongly nonadiabatic part of the evolution has ended.

During the inflationary phase, for modes satisfying
$k^2\gg |m_{\rm eff}^2|$, the adiabatic basis reduces at leading order to
the plane-wave basis,
\begin{equation}
u_k^{(4)}(\eta)\simeq\frac{e^{-ik\eta}}{\sqrt{2k}}.
\end{equation}
Equation~(\ref{eq:IV-final-alpha-beta}) then reduces to
\begin{align}
\alpha^{(4)}_{k,f}
&=
\sqrt{\frac{k}{2}}\,e^{ik\eta_f}
\left[
\nu_k(\eta_f)+\frac{i}{k}\nu'_k(\eta_f)
\right],\nonumber\\
 \beta^{(4)}_{k,f}
&=
\sqrt{\frac{k}{2}}\,e^{-ik\eta_f}
\left[
\nu_k(\eta_f)-\frac{i}{k}\nu'_k(\eta_f)
\right].
\label{eq:IV-plane-wave-projection}
\end{align}
These relations provide a direct matching of the exact evolved mode to the
late-time particle basis.

\begin{figure}[h]
{\includegraphics[width=8cm]{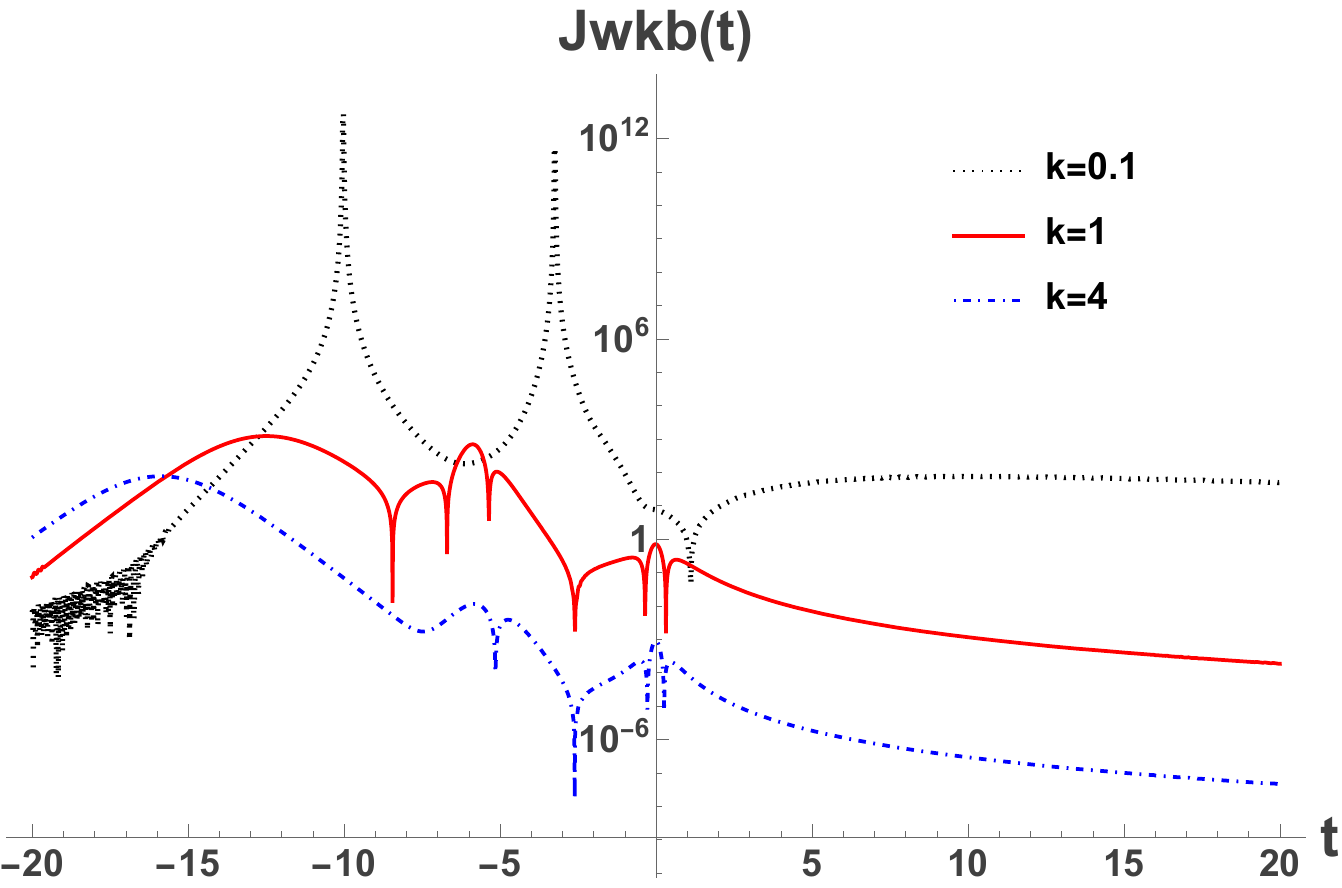}}
\caption{The adiabaticity parameter ${\rm Jwkb}(t)$ defined in
Eq.~(\ref{eq:IV-Jwkb}) for mLQC-I in the hybrid approach, shown for
$k/m_P=0.1$, $1$, and $4$.  Primes in Eq.~(\ref{eq:IV-Jwkb}) denote
derivatives with respect to conformal time, while the result is displayed
here as a function of cosmic time $t$.  The figure diagnoses
nonadiabatic evolution for these finite modes; the sharp behavior near
turning points indicates the breakdown of the WKB approximation and should
not be extrapolated to the strict $k\rightarrow\infty$ limit.}
\label{fig10}
\end{figure}

The amount of particle production is controlled dynamically by the violation
of adiabaticity.  A useful diagnostic is
\begin{equation}
{\rm Jwkb}(t)
=
\left|
\frac{3{\omega_k'}^2}{4\omega_k^4}
-\frac{\omega_k''}{2\omega_k^3}
\right|,
\label{eq:IV-Jwkb}
\end{equation}
where the primes denote derivatives with respect to conformal time $\eta$;
in Fig.~\ref{fig10} the resulting quantity is displayed as a function of
cosmic time $t$.  The quantity ${\rm Jwkb}$ can become large for finite and
intermediate wavenumbers during the pre-bounce evolution, with stronger
violations for smaller $k$.  In particular, it becomes singular at a turning
point where $\omega_k^2=0$; this signals the breakdown of the WKB expansion
there and should not be interpreted as a divergent particle number.
Nevertheless, the strong violation of adiabaticity shows that the pre-bounce
evolution can generate a nonzero $\beta_{k,f}^{(4)}$ and particle creation
for finite and intermediate wavenumbers, even though the ultraviolet tail
remains adiabatic.

\subsection{Ultraviolet regularity and the inflationary phase}
\label{sec:IV-UV-regularity}

The ultraviolet behavior follows directly from the adiabatic matching above.
The initial de Sitter state satisfies
Eq.~(\ref{eq:IV-initial-alpha-beta}), and the effective mass
$m_{\rm eff}^2(\eta)$ is smooth throughout the evolution considered here.
Exact linear evolution on this smooth background therefore preserves
the ultraviolet adiabatic order.
Consequently,
\begin{equation}
\beta_{k,f}^{(4)}
=
{\cal O}(k^{-5}),
\quad k\rightarrow\infty,
\label{eq:IV-beta-UV-falloff}
\end{equation}
whereas finite and intermediate modes can have nonzero
$\beta_{k,f}^{(4)}$.  Equation~(\ref{eq:IV-beta-UV-falloff}) is the
ultraviolet result relevant to particle creation; it is qualitatively
different from the $k$-independent negative-frequency coefficient obtained
by extrapolating the leading finite-$k$ Airy connection.

The same conclusion can be expressed directly in terms of the renormalized
energy density.  Define
\begin{equation}
q_k\equiv\frac{\nu_k}{a},
\quad
s(\eta)=m_{\rm eff}^2(\eta)
=
U_{\rm eff}(\eta)-\frac{a''}{a}.
\label{eq:IV-s-Ueff}
\end{equation}
Adiabatic regularization
\cite{Parker:1969au,PhysRevD.9.341,Parker_Toms_2009} gives
\footnote{In this work we focus on the ultraviolet behavior of the
momentum integral. A possible infrared divergence as $k\to0$ can be
regulated, following Refs.~\cite{Agullo:2012fc,Agullo:2013ai}, by
introducing a minimum comoving wavenumber $k_{\min}$ such that the
corresponding longest physical wavelength today, $a_0/k_{\min}$, is
comparable to the size of the observable universe. This infrared
cutoff does not affect the ultraviolet analysis considered here.}
\begin{equation}
\rho_{\rm ren}(\eta)
=
\frac{\hbar}{2a^2}
\int\frac{d^3k}{(2\pi)^3}
\left[
I_k^{\rm bare}-C^{(Q)}(k,\eta)
\right],
\label{eq:IV-rho-ren}
\end{equation}
where
\begin{equation}
I_k^{\rm bare}
=
|q_k'|^2+(k^2+U_{\rm eff})|q_k|^2,
\label{eq:IV-Ibare}
\end{equation}
and the fourth-order subtraction term is
\begin{align}
C^{(Q)}(k,\eta)
=&\frac{k}{a^2}
+\frac{a'^2+a^2U_{\rm eff}}{2a^4k}
\nonumber\\
&+\frac{1}{8a^5k^3}
\Big[
-a^3U_{\rm eff}^2
+2a^2a'U_{\rm eff}'
+4a'^2a''\nb\\
&
+a\big(
a''^2-2a'(U_{\rm eff}a'+a''')
\big)
\Big].
\label{eq:IV-CQ}
\end{align}
Since the evolved state remains of fourth adiabatic order,
\begin{equation}
I_k^{\rm bare}-C^{(Q)}(k,\eta)
=
{\cal O}(k^{-5}),
\label{eq:IV-UV-remainder}
\end{equation}
at the bounce and during the subsequent inflationary phase.  Because the
three-dimensional momentum measure contributes a factor $k^2$, the
ultraviolet part of Eq.~(\ref{eq:IV-rho-ren}) behaves as
$\int^\infty dk\,k^{-3}$ or faster and is therefore convergent.

Therefore, particle creation and ultraviolet regularity describe two different
parts of the same evolution.  The strongly time-dependent pre-bounce
background can generate a nonzero $\beta_{k,f}^{(4)}$ for finite and
intermediate modes, and this finite-mode particle population can persist
into the inflationary phase after the evolution becomes adiabatic again.
At the same time, Eq.~(\ref{eq:IV-beta-UV-falloff}) shows that
$\beta_{k,f}^{(4)}\rightarrow0$ in the ultraviolet with the falloff required
by fourth-order adiabaticity, so no ultraviolet divergence is generated.
The constant negative-frequency coefficient obtained by extrapolating the
leading finite-$k$ Airy connection is therefore not a physical ultraviolet
particle population but an artifact of applying that connection outside its
asymptotic domain.

\section{Initial Conditions Imposed at the Bounce,  Ultraviolet Regularity and Particle Creation}
\lb{SecV}

We now consider the second prescription introduced in Sec.~I, in which the
perturbations are placed in a fourth-order adiabatic state directly at the
quantum bounce \cite{Agullo:2012fc,Agullo:2012sh,Agullo:2013ai}.  In contrast
with the remote de Sitter prescription of Sec.~\ref{SecIV}, no matching to the
pre-bounce branch is required to define this state.  The initial data are
specified locally at the bounce through the fourth-order adiabatic mode and
its first derivative.  We first show explicitly that these data have the
ultraviolet structure required for adiabatic renormalization, and then evolve
the exact MS equation into the post-bounce and inflationary
phases to determine the resulting particle content.

\subsection{Fourth-order adiabatic initial state at the bounce and ultraviolet regularity}
\label{sec:V-bounce-state}

As in Sec.~\ref{sec:IV-dS-adiabatic}, we write
\begin{equation}
s(\eta)\equiv m_{\rm eff}^2(\eta)
=U_{\rm eff}(\eta)-\frac{a''}{a}.
\label{eq:V-seff}
\end{equation}
For the mode equation
\begin{equation}
\nu_k''+\left(k^2+s\right)\nu_k=0,
\label{eq:V-mode}
\end{equation}
let $W_k^{(4)}$ denote the fourth-order adiabatic frequency.  The
ultraviolet expansion of this frequency is
\begin{equation}
W_k^{(4)}
=
k+\frac{s}{2k}
-\frac{s^2+s''}{8k^3}
+{\cal O}(k^{-5}),
\label{eq:V-W4}
\end{equation}
with
\begin{equation}
W_k^{(4)\prime}
=
\frac{s'}{2k}
-\frac{2ss'+s'''}{8k^3}
+{\cal O}(k^{-5}).
\label{eq:V-W4-prime}
\end{equation}
Equations~(\ref{eq:V-W4}) and (\ref{eq:V-W4-prime}) are used below only
to establish the large-$k$ structure of the state.  The state itself is
defined by the fourth-order adiabatic mode and its first derivative at the
bounce.

The fourth-order adiabatic state imposed at the bounce,
$\eta=\eta_B$, is defined by
\bqn
\nu_k(\eta_B)
&=&
\frac{1}{\sqrt{2W_{k,B}^{(4)}}},
\nb\\
\nu_k'(\eta_B)
&=&
\left[
-\frac{W_{k,B}^{(4)\prime}}{2W_{k,B}^{(4)}}
-iW_{k,B}^{(4)}
\right]\nu_k(\eta_B),
\label{eq:V-initialdata}
\eqn
where
\begin{equation}
W_{k,B}^{(4)}\equiv W_k^{(4)}(\eta_B).
\end{equation}
These two conditions specify both the mode and its canonical momentum and
therefore define the state through fourth adiabatic order.  In particular,
with respect to the fourth-order adiabatic basis chosen at the bounce, the
initial Bogoliubov coefficients are
\begin{equation}
\alpha_{k,B}^{(4)}=1,
\quad
\beta_{k,B}^{(4)}=0.
\label{eq:V-bounce-bogoliubov}
\end{equation}
Thus the state contains no negative-frequency component at the instant at
which it is imposed, with respect to this adiabatic basis.

We next verify explicitly that the state is ultraviolet regular.  With
\begin{equation}
q_k\equiv\frac{\nu_k}{a},
\end{equation}
the bare energy-density integrand is
\begin{equation}
I_k^{\rm bare}
=
|q_k'|^2+\left(k^2+U_{\rm eff}\right)|q_k|^2.
\label{eq:V-Ibare-def}
\end{equation}
Using Eq.~(\ref{eq:V-initialdata}), one finds at the bounce
\bqn
&& |q_k'|^2
=
\frac{1}{a^2}\frac{1}{2W_k^{(4)}}
\left[
\left(W_k^{(4)}\right)^2
+\left(
\frac{W_k^{(4)\prime}}{2W_k^{(4)}}+\frac{a'}{a}
\right)^2
\right],
\nb\\
&& \left(k^2+U_{\rm eff}\right)|q_k|^2
=
\frac{k^2+U_{\rm eff}}{a^2}\frac{1}{2W_k^{(4)}},
\label{eq:V-q-pieces}
\eqn
where all background quantities in Eq.~(\ref{eq:V-q-pieces}) are evaluated
at $\eta_B$.  Hence
\bqn
I_k^{\rm bare}
&=&
\frac{1}{a^2}\frac{1}{2W_k^{(4)}}
\Bigg[
\left(W_k^{(4)}\right)^2
+\left(
\frac{W_k^{(4)\prime}}{2W_k^{(4)}}+\frac{a'}{a}
\right)^2
\nb\\
&&\hspace{2.0cm}
+k^2+U_{\rm eff}
\Bigg].
\label{eq:V-Ibare-W}
\eqn

Substituting Eqs.~(\ref{eq:V-seff})--(\ref{eq:V-W4-prime}) into
Eq.~(\ref{eq:V-Ibare-W}) and expanding in inverse powers of $k$, the terms
of orders $k$, $k^{-1}$, and $k^{-3}$ reproduce precisely the fourth-order
subtraction term $C^{(Q)}(k,\eta_B)$ given in
Eq.~(\ref{eq:IV-CQ}).  Therefore,
\begin{equation}
I_k^{\rm bare}-C^{(Q)}(k,\eta_B)
=
{\cal O}(k^{-5}).
\label{eq:V-bounce-finite}
\end{equation}
Since the momentum-space measure contributes a factor $k^2$, the ultraviolet
part of the renormalized energy-density integral behaves as
$\int^\infty dk\,k^{-3}$ or faster and is convergent.  Notice that this
statement concerns ultraviolet regularity; it does not imply that the finite
renormalized energy density at the bounce vanishes.

The ultraviolet conclusion follows directly from the adiabatic initial data
and does not require the Airy/UAA coefficients $\widetilde a_k$ and
$\widetilde b_k$.  The UAA remains useful for describing finite- and
intermediate-wavenumber mode evolution, but the ultraviolet state at the
bounce is completely specified by Eq.~(\ref{eq:V-initialdata}).

\subsection{Exact post-bounce evolution and particle creation}
\label{sec:V-postbounce}

Starting from Eq.~(\ref{eq:V-initialdata}), we evolve the exact
MS equation forward from the bounce.  As in Sec.~IV, particle
creation is defined only after the mode has entered a regime in which a
positive/negative-frequency adiabatic decomposition is meaningful.

Let $\eta_f>\eta_B$ be such a time and let $u_{k,f}^{(4)}$ denote the
fourth-order positive-frequency adiabatic mode constructed from the exact
background at $\eta_f$.  The evolved mode can be written as
\begin{equation}
\nu_k(\eta_f)
=
\alpha_{k,f}^{B}\,u_{k,f}^{(4)}
+
\beta_{k,f}^{B}\,u_{k,f}^{(4)*},
\label{eq:V-late-decomp}
\end{equation}
where the superscript $B$ labels the state selected at the bounce; the
positive- and negative-frequency basis at $\eta_f$ is the fourth-order
adiabatic basis.  The corresponding Bogoliubov coefficients are
\begin{align}
\alpha_{k,f}^{B}
&=
i\left[
u_{k,f}^{(4)*}\nu_k'(\eta_f)
-u_{k,f}^{(4)*\prime}\nu_k(\eta_f)
\right],
\nonumber\\
\beta_{k,f}^{B}
&=
-i\left[
u_{k,f}^{(4)}\nu_k'(\eta_f)
-u_{k,f}^{(4)\prime}\nu_k(\eta_f)
\right],
\label{eq:V-final-alpha-beta}
\end{align}
and satisfy
\begin{equation}
\left|\alpha_{k,f}^{B}\right|^2-\left|\beta_{k,f}^{B}\right|^2=1.
\label{eq:V-final-Wronskian}
\end{equation}
The particle number associated with this late-time adiabatic basis is
\begin{equation}
N_k^{B}=\left|\beta_{k,f}^{B}\right|^2.
\label{eq:V-particle-number}
\end{equation}

Although $\beta_{k,B}^{(4)}=0$ at the bounce by construction, the exact
post-bounce evolution need not preserve this condition for finite $k$.
Whenever the background varies nonadiabatically, positive- and
negative-frequency components can mix and a nonzero
$\beta_{k,f}^{B}$ can be generated.  The adiabaticity diagnostic introduced
in Eq.~(\ref{eq:IV-Jwkb}) and displayed in Fig.~\ref{fig10} becomes small
again in the post-bounce branch for the representative modes shown there.
This indicates that additional particle creation is suppressed once the
evolution has returned to an adiabatic regime.  As emphasized in Sec.~IV,
the sharp behavior of this diagnostic near a turning point reflects the
breakdown of the WKB expansion there and is not itself a particle number.

During the inflationary phase, for modes satisfying
$k^2\gg |m_{\rm eff}^2|$, the adiabatic basis reduces at leading order to
the plane-wave basis and
\begin{equation}
\nu_k(\eta)
\simeq
\frac{1}{\sqrt{2k}}
\left(
\alpha_{k, f}^{B}e^{-ik\eta}
+
\beta_{k, f}^{B}e^{ik\eta}
\right).
\label{eq:V-inflation-bog}
\end{equation}
Thus a finite-wavenumber particle population created after the bounce can
persist into the inflationary phase even after the subsequent evolution has
become adiabatic.

\subsection{Preservation of ultraviolet regularity}
\label{sec:V-UV-preservation}

The finite-$k$ particle production described above is distinct from the
ultraviolet behavior of the state.  Because the initial data
(\ref{eq:V-initialdata}) define a fourth-order adiabatic state and the
effective mass $m_{\rm eff}^2(\eta)$ is smooth in the evolution considered
here, exact linear evolution preserves the ultraviolet adiabatic structure.
Consequently, at any later time for which the effective description remains
valid,
\begin{equation}
I_k^{\rm bare}-C^{(Q)}(k,\eta)
=
{\cal O}(k^{-5}),
\label{eq:V-postbounce-finite}
\end{equation}
and the ultraviolet part of the renormalized energy density remains finite.

Correspondingly, when the evolved mode is projected onto a
fourth-order adiabatic basis in a late-time adiabatic regime, its
negative-frequency coefficient must vanish in the ultraviolet, 
\begin{equation}
\beta_{k,f}^{B}\longrightarrow0,
\quad k\rightarrow\infty,
\label{eq:V-beta-UV}
\end{equation}
with the falloff required by fourth-order adiabaticity.  Therefore
$\beta_{k,f}^{B}$ may be nonzero for finite and intermediate wavenumbers,
but it cannot approach a nonzero constant in the ultraviolet.  The
ultraviolet convergence itself is established directly by
Eq.~(\ref{eq:V-postbounce-finite}).  Finite-wavenumber particle creation
is therefore fully compatible with fourth-order ultraviolet regularity.

The two initial-state prescriptions considered in Secs.~IV and V therefore
share the same ultraviolet consistency property but differ in where and how
the state is selected.  In Sec.~IV the state is fixed by the asymptotic de
Sitter symmetry in the remote contracting branch and is then evolved through
the bounce.  Here the state is imposed locally at the bounce by the
fourth-order adiabatic conditions (\ref{eq:V-initialdata}).  Their physically
interesting differences are consequently encoded in the finite- and
intermediate-wavenumber mode evolution and in the resulting particle spectra,
rather than in the ultraviolet finiteness of the renormalized energy density.

\section{Conclusions and Outlook}

In this work we have developed a high-accuracy description of cosmological
perturbations in mLQC-I and studied the associated ultraviolet regularity and
gravitational particle creation.  The detailed construction of the effective
mass function and of the mode functions in Secs.~II and III has been carried
out in the hybrid approach.  In this case the evolution naturally separates
into an asymptotic pre-de Sitter branch, a pre-inflationary region containing
the quantum bounce, and the subsequent inflationary phase.  The mode functions
are known analytically in the two asymptotic regimes, while the
one-turning-point problem in the intermediate region can be treated by the
UAA method.

A main goal of the present work has been to obtain approximate mode functions
with sufficiently high accuracy, including both amplitude and phase, for
future calculations of observables such as the primordial power spectrum and
non-Gaussian correlation functions.  The leading Airy approximation captures
the qualitative behavior through the turning point, but its accuracy is not
uniform over the relevant range of wavenumbers.  We therefore constructed the
complete third-order UAA, keeping the general pre-de Sitter index and the
finite parts of the asymptotic phase integrals.  For the representative modes
$k=5m_P$ and $10m_P$, the normalized complex-mode RMS error is reduced to
below one percent, while the lower-wavenumber modes are more demanding.  In
particular, the observationally relevant interval begins near $k\simeq m_P$,
above the crossover scale, so that the complete third-order UAA provides the
preferred analytic seed throughout this range, with the defect correction
supplying the additional accuracy required for precision applications.

To obtain a uniformly accurate representation also in the more difficult
low-$k$ regime, we supplemented the UAA with a defect correction.  The
preferred seed is chosen adaptively: the first-order Airy approximation is
used for $k<k_c$, while the complete third-order approximation is used for
$k\geq k_c$, with
\begin{equation}
k_c\simeq0.5457\,m_P .
\end{equation}
The residual of the chosen seed is then propagated with the exact
MS operator.  Over the range
$0.1\leq k/m_P\leq1$, the resulting defect-corrected mode functions agree
with the direct numerical solutions at the level of
\bq
\lb{eq6.2}
{\cal E}_{\nu}^{\rm corr} = {\cal O}\left(10^{-5}\%\right). 
\eq
This accuracy should not be interpreted as a new
formal UAA truncation bound; at this level it is limited mainly by the
numerical residual propagation, interpolation, and projection.  Its practical
importance is that the construction provides a highly accurate semi-analytic
mode function while retaining the analytic organization of the UAA.

The corresponding finite-wavenumber connection has been written in transfer
matrix form.  This representation keeps the amplitudes and relative phases of
the positive- and negative-frequency components and can be corrected
systematically by the defect matrix.  These features are particularly useful
for future calculations.  The primordial power spectrum depends directly on
the late-time mode amplitude, whereas higher-order correlation functions,
including those characterizing primordial non-Gaussianity, also depend
sensitively on the relative phases and on the time dependence of the mode
functions inside the interaction integrals.  The present construction
therefore provides a controlled starting point for precision studies of both
the power spectrum and non-Gaussianity in mLQC-I.  The accuracy of the
resulting observables should be assessed directly in those calculations.

A second central result concerns the distinction between finite-wavenumber
matching and the strict ultraviolet limit.  The connection formulas of
Sec.~III are derived in a small-$y$ overlap region, with $y=k\eta\ll1$,
whereas at any fixed nonzero conformal time the ultraviolet limit
$k\rightarrow\infty$ implies $y\rightarrow\infty$.  These are different
asymptotic limits and cannot be interchanged.  Consequently, a finite-order
Airy connection may approximate the full finite-$k$ mode very well while
still failing to resolve the much smaller ultraviolet negative-frequency
component.  In particular, a $k$-independent Bogoliubov coefficient obtained
by extrapolating the leading Airy formula to $k\rightarrow\infty$ does not
represent the ultraviolet behavior of the exact state.

For the de Sitter vacuum imposed in the remote contracting branch, we have therefore treated the ultraviolet sector separately
by matching the large-$k$ Hankel expansion to the fourth-order
adiabatic expansion.   The
large-$k$ de Sitter mode and its first derivative agree with the fourth-order
adiabatic mode and its first derivative through the order required for
energy-momentum tensor renormalization.  Because the effective mass is smooth, exact
evolution preserves this ultraviolet adiabatic structure.  At the bounce and
during the subsequent inflationary phase one consequently has
\begin{equation}
I_k^{\rm bare}-C^{(Q)}(k,\eta)
=
{\cal O}\left(k^{-5}\right),
\end{equation}
so that the ultraviolet part of the renormalized energy density is finite.
At the same time, the strongly nonadiabatic evolution of finite and
intermediate modes can mix positive- and negative-frequency components and
produce particles.  Thus finite-mode particle creation and ultraviolet
regularity coexist naturally.

We have also considered a fourth-order adiabatic state imposed directly at
the quantum bounce.  This state is ultraviolet regular by construction, and
its exact post-bounce evolution preserves the same fourth-order ultraviolet
structure.  Although the late-time negative-frequency coefficient can be
nonzero for finite $k$, it must vanish in the ultraviolet with the falloff
required by fourth-order adiabaticity.  The two initial-state prescriptions
studied here therefore differ primarily in their finite- and
intermediate-wavenumber evolution and in the resulting particle spectra,
rather than in the ultraviolet finiteness of the renormalized energy density.

We have focused in this work on ultraviolet regularity.  A possible
infrared divergence as $k\to0$ is a separate issue and can be regulated,
following Refs.~\cite{Agullo:2012fc,Agullo:2013ai}, by introducing a
minimum comoving wavenumber $k_{\min}$ such that the corresponding
longest physical wavelength today, $a_0/k_{\min}$, is comparable to the
size of the observable universe.  This infrared cutoff does not affect
any of the ultraviolet conclusions obtained above.

The quantitative results of Secs.~II and III, including the effective mass,
the precise turning-point structure, the crossover scale $k_c$, and the
numerical accuracy of the higher-order and defect-corrected mode functions,
are specific to the hybrid approach.  The ultraviolet-regularity argument and
the framework for particle creation, however, are more general.  In the
dressed-metric approach, in the large-$k$ regime relevant to the present
ultraviolet analysis, the mode equation likewise possesses a single simple
turning point, so that the corresponding uniform approximation is again of
Airy type \cite{Pan:2025ppa}.  The same distinction between finite-$k$ Airy matching and the
strict large-$k$ adiabatic limit therefore applies.  A constant
negative-frequency coefficient obtained from a leading Airy connection
cannot by itself establish an ultraviolet pathology, and finite-wavenumber
particle creation remains compatible with ultraviolet regularity.  The
quantitative particle spectrum and the corresponding higher-order mode
functions for any given wavenumber in the dressed-metric approach must nevertheless be recalculated
with its own effective mass.

The present results thus provide both a precision tool and a conceptual
framework.  On the technical side, the higher-order UAA together with the
adaptive defect correction yields complex mode functions of sufficient
accuracy for future precision calculations of the primordial power spectrum
and non-Gaussian correlations.  On the conceptual side, the analysis shows
that the approximation used to describe finite modes and the asymptotic
expansion required to control ultraviolet renormalization must be kept
distinct.  This separation will be important in extending the present study
to detailed predictions for primordial observables and to corresponding
analyses in the dressed-metric approach.

\begin{acknowledgments}

We are grateful to Prof. Ivan Agullo for valuable discussions
and insightful comments. J.S. and R.P. acknowledge the
generous support of Baylor University through the Department
of Physics and Astronomy graduate program. A.W. is partially
supported by the U.S. National Science Foundation under
Grant No. PHY-2308845. We also acknowledge the use of ChatGPT as a computational
aid in performing some mathematical calculations.
 
\end{acknowledgments}

\appendix
\section{Fundamental-Matrix Representation of the Defect Correction}
\label{app:defect-propagator}

This appendix gives a compact matrix representation of the defect correction introduced in Secs.~\ref{sec:defect} and~\ref{sec:IIIE-matching}.  The result is equivalent to solving the two basis defect equations in Eq.~(\ref{eq:IIIE-basis-defect-eq}), but is useful for analytical manipulations and for independent numerical implementations.

For either seed order $s=1$ or $3$, define the two-component defect vector for the incoming basis state $j=a,b$ by
\begin{equation}
 {\bf X}_{j,k}^{(s)}(y)\equiv
 \begin{pmatrix}
 \delta\phi_{j,k}^{(s)}(y)\\[1mm]
 \partial_y\delta\phi_{j,k}^{(s)}(y)
 \end{pmatrix}.
 \label{eq:app-Xdef}
\end{equation}
Equation~(\ref{eq:IIIE-basis-defect-eq}) can then be written as the first-order system
\begin{equation}
 \frac{d}{dy}{\bf X}_{j,k}^{(s)}(y)
 =A_k(y){\bf X}_{j,k}^{(s)}(y)-{\bf e}_2\,r_{j,k}^{(s)}(y),
 \label{eq:app-firstorder}
\end{equation}
where
\begin{equation}
 A_k(y)=
 \begin{pmatrix}
 0&1\\
 Q_k(y)&0
 \end{pmatrix},
\quad
 {\bf e}_2=\begin{pmatrix}0\\1\end{pmatrix}.
 \label{eq:app-Ae2}
\end{equation}
Let $U_k(y,u)$ be the exact fundamental matrix of the homogeneous system,
\begin{equation}
 \partial_y U_k(y,u)=A_k(y)U_k(y,u),
\quad
 U_k(u,u)=I.
 \label{eq:app-propagator}
\end{equation}
Variation of constants gives
\bqn
 {\bf X}_{j,k}^{(s)}(y_f)
 &=&U_k(y_f,y_i){\bf X}_{j,k}^{(s)}(y_i)\nb\\
 && 
 -\int_{y_i}^{y_f}U_k(y_f,u){\bf e}_2\,
 r_{j,k}^{(s)}(u)\,du.
 \label{eq:app-Xsolution}
\eqn 

It is convenient to collect the two incoming basis states into matrices.  Define
\begin{equation}
 D_{k,i}^{(s)}\equiv
 \begin{pmatrix}
 \delta\phi_{a,k}^{(s)}(y_i) & \delta\phi_{b,k}^{(s)}(y_i)\\[1mm]
 \partial_y\delta\phi_{a,k}^{(s)}(y_i) &
 \partial_y\delta\phi_{b,k}^{(s)}(y_i)
 \end{pmatrix},
 \label{eq:app-Di}
\end{equation}
and the residual row vector
\begin{equation}
 {\bf r}_{k}^{(s)}(u)\equiv
 \begin{pmatrix}
 r_{a,k}^{(s)}(u) & r_{b,k}^{(s)}(u)
 \end{pmatrix}.
 \label{eq:app-rrow}
\end{equation}
The matrix whose columns are the two late-time defect vectors is therefore
\begin{equation}
 X_{k,f}^{(s)}=
 U_k(y_f,y_i)D_{k,i}^{(s)}
 -\int_{y_i}^{y_f}
 U_k(y_f,u){\bf e}_2\,{\bf r}_{k}^{(s)}(u)\,du.
 \label{eq:app-Xmatrix}
\end{equation}

In the oscillatory region, the map from a mode and its $y$ derivative to the positive- and negative-frequency coefficients is
\begin{equation}
 {\cal P}_k(y_f)
 =\sqrt{\frac{k}{2}}
 \begin{pmatrix}
 e^{iy_f} & i e^{iy_f}\\
 e^{-iy_f} & -i e^{-iy_f}
 \end{pmatrix}.
 \label{eq:app-Pmatrix}
\end{equation}
Consequently, the correction to the seed transfer matrix has the compact propagator form
\bqn 
 \Delta T_k^{(s)}&=&
 {\cal P}_k(y_f)
 \left[
 U_k(y_f,y_i)D_{k,i}^{(s)}\right.\nb\\
 &&\left.
 -\int_{y_i}^{y_f}
 U_k(y_f,u){\bf e}_2\,{\bf r}_{k}^{(s)}(u)\,du
 \right].
 \label{eq:app-DeltaT-propagator}
\eqn 
This expression separates naturally into an initial-data mismatch and a propagated residual,
\begin{equation}
 \Delta T_k^{(s)}
 =\Delta T_{k,{\rm IC}}^{(s)}
 +\Delta T_{k,{\rm res}}^{(s)},
 \label{eq:app-DeltaT-split}
\end{equation}
with
\begin{align}
 \Delta T_{k,{\rm IC}}^{(s)}
 &= {\cal P}_k(y_f)U_k(y_f,y_i)D_{k,i}^{(s)},
 \nonumber\\
 \Delta T_{k,{\rm res}}^{(s)}
 &=-{\cal P}_k(y_f)
 \int_{y_i}^{y_f}
 U_k(y_f,u){\bf e}_2\,{\bf r}_{k}^{(s)}(u)\,du.
 \label{eq:app-DeltaT-pieces}
\end{align}
The first term corrects the finite-$y_i$ mismatch between the analytic pre-de Sitter basis and the chosen UAA seed, while the second propagates the accumulated residual of that seed through the exact mode operator.

Finally, the Wronskian provides a useful numerical consistency check.  The pre-de Sitter and inflationary coefficients satisfy, respectively,
\begin{equation}
\left|a_k\right|^2-\left|b_k\right|^2= \frac{\pi}{4k},
\quad
 \left|\alpha_k\right|^2-\left|\beta_k\right|^2=1.
 \label{eq:app-Wronskians}
\end{equation}
Hence the exact corrected transfer matrix must obey
\begin{equation}
 \left(T_k^{\rm corr}\right)^\dagger
 \sigma_3 T_k^{\rm corr}
 =\frac{4k}{\pi}\,\sigma_3,
\quad
 \sigma_3=\begin{pmatrix}1&0\\0&-1\end{pmatrix}.
 \label{eq:app-pseudounitary}
\end{equation}
Violation of Eq.~(\ref{eq:app-pseudounitary}) therefore directly measures numerical error in the corrected propagation and late-time projection.

\bibliography{BibNVar}

@article{AtacamaCosmologyTelescope:2025nti,
    author = "Calabrese, Erminia and others",
    collaboration = "Atacama Cosmology Telescope",
    title = "{The Atacama Cosmology Telescope: DR6 constraints on extended cosmological models}",
    eprint = "2503.14454",
    archivePrefix = "arXiv",
    primaryClass = "astro-ph.CO",
    reportNumber = "FERMILAB-PUB-25-0157-PPD",
    doi = "10.1088/1475-7516/2025/11/063",
    journal = "JCAP",
    volume = "11",
    pages = "063",
    year = "2025"
}

@article{AtacamaCosmologyTelescope:2025blo,
    author = "Louis, Thibaut and others",
    collaboration = "Atacama Cosmology Telescope",
    title = "{The Atacama Cosmology Telescope: DR6 power spectra, likelihoods and {\ensuremath{\Lambda}}CDM parameters}",
    eprint = "2503.14452",
    archivePrefix = "arXiv",
    primaryClass = "astro-ph.CO",
    reportNumber = "FERMILAB-PUB-25-0071-PPD",
    doi = "10.1088/1475-7516/2025/11/062",
    journal = "JCAP",
    volume = "11",
    pages = "062",
    year = "2025"
}

@book{CMB-S4:2016ple,
    author = "Abazajian, Kevork N. and others",
    collaboration = "CMB-S4",
    title = "{CMB-S4 Science Book, First Edition}",
    eprint = "1610.02743",
    archivePrefix = "arXiv",
    primaryClass = "astro-ph.CO",
    reportNumber = "FERMILAB-FN-1024-A-AE",
    doi = "10.2172/1352047",
    month = "10",
    year = "2016"
}

@article{Han:2021cwb,
    author = "Han, Muxin and Liu, Hongguang",
    title = "{Loop quantum gravity on dynamical lattice and improved cosmological effective dynamics with inflaton}",
    eprint = "2101.07659",
    archivePrefix = "arXiv",
    primaryClass = "gr-qc",
    doi = "10.1103/PhysRevD.104.024011",
    journal = "Phys. Rev. D",
    volume = "104",
    number = "2",
    pages = "024011",
    year = "2021"
}

@article{Assanioussi:2018hee,
    author = "Assanioussi, Mehdi and Dapor, Andrea and Liegener, Klaus and Paw{\l}owski, Tomasz",
    title = "{Emergent de Sitter Epoch of the Quantum Cosmos from Loop Quantum Cosmology}",
    eprint = "1801.00768",
    archivePrefix = "arXiv",
    primaryClass = "gr-qc",
    doi = "10.1103/PhysRevLett.121.081303",
    journal = "Phys. Rev. Lett.",
    volume = "121",
    number = "8",
    pages = "081303",
    year = "2018"
}

@article{Dapor:2017rwv,
    author = "Dapor, Andrea and Liegener, Klaus",
    title = "{Cosmological Effective Hamiltonian from full Loop Quantum Gravity Dynamics}",
    eprint = "1706.09833",
    archivePrefix = "arXiv",
    primaryClass = "gr-qc",
    doi = "10.1016/j.physletb.2018.09.005",
    journal = "Phys. Lett. B",
    volume = "785",
    pages = "506--510",
    year = "2018"
}

@article{Giesel:2026fsp,
    author = "Giesel, Kristina and Guill{\'e}n, Almudena and Mena Marug{\'a}n, Guillermo A. and Ricci, Leonardo",
    title = "{Robustness of the primordial power spectrum in hybrid loop quantum cosmology to approximations near the bounce}",
    eprint = "2608.12181",
    archivePrefix = "arXiv",
    primaryClass = "gr-qc",
    month = "8",
    year = "2026"
}

@article{ElizagaNavascues:2019itm,
    author = "Elizaga Navascu\'es, Beatriz and Marug\'an, Guillermo A. Mena and Thiemann, Thomas",
    title = "{Hamiltonian diagonalization in hybrid quantum cosmology}",
    eprint = "1903.05695",
    archivePrefix = "arXiv",
    primaryClass = "gr-qc",
    doi = "10.1088/1361-6382/ab32af",
    journal = "Class. Quant. Grav.",
    volume = "36",
    number = "18",
    pages = "185010",
    year = "2019"
}

@article{Gan:2025uvt,
    author = "Gan, Wen-Cong and Graef, Leila L. and O. Ramos, Rudnei and S. Vicente, Gustavo and Wang, Anzhong",
    title = "{Quantum Damping of Cosmological Shear: A New Prediction from Loop Quantum Cosmologies}",
    eprint = "2510.14021",
    archivePrefix = "arXiv",
    primaryClass = "gr-qc",
    doi = "10.1103/f8tr-bq61",
    journal = "Phys. Rev. Lett.",
    volume = "136",
    number = "25",
    pages = "251501",
    year = "2026"
}

@article{Gan:2026ivx,
    author = "Gan, Wen-Cong and Graef, Leila L. and O. Ramos, Rudnei and S. Vicente, Gustavo and Wang, Anzhong",
    title = "{Genericness of quantum damping of cosmological shear in modified loop quantum cosmology}",
    eprint = "2604.08912",
    archivePrefix = "arXiv",
    primaryClass = "gr-qc",
    month = "4",
    year = "2026"
}

@article{Graef:2020gpp,
    author = "Graef, L. L. and Ramos, Rudnei O. and Vicente, G. S.",
    title = "{Gravitational particle production in loop quantum cosmology}",
    eprint = "2007.02395",
    archivePrefix = "arXiv",
    primaryClass = "gr-qc",
    doi = "10.1103/PhysRevD.102.043518",
    journal = "Phys. Rev. D",
    volume = "102",
    number = "4",
    pages = "043518",
    year = "2020"
}

@article{Vicente:2022gpp,
    author = "Vicente, G. S. and Ramos, Rudnei O. and Graef, L. L.",
    title = "{Gravitational particle production and the validity of
    effective descriptions in loop quantum cosmology}",
    eprint = "2207.00435",
    archivePrefix = "arXiv",
    primaryClass = "gr-qc",
    doi = "10.1103/PhysRevD.106.043518",
    journal = "Phys. Rev. D",
    volume = "106",
    number = "4",
    pages = "043518",
    year = "2022"
}

@article{Pan:2025ppa,
    author = "Pan, Rui and Saeed, Jamal and Wang, Anzhong",
    title = "{Stable initial conditions and analytically approximate solutions of cosmological perturbations in a modified loop quantum cosmology}",
    eprint = "2505.14813",
    archivePrefix = "arXiv",
    primaryClass = "gr-qc",
    doi = "10.1088/1475-7516/2026/06/030",
    journal = "JCAP",
    volume = "06",
    pages = "030",
    year = "2026"
}

@book{Parker_Toms_2009, place={Cambridge}, series={Cambridge Monographs on Mathematical Physics}, title={Quantum Field Theory in Curved Spacetime: Quantized Fields and Gravity, $~$}, publisher={Cambridge University Press}, author={Parker, Leonard and Toms, David}, year={2009}, collection={Cambridge Monographs on Mathematical Physics}}

@article{Parker:1969au,
    author = "Parker, Leonard",
    title = "{Quantized fields and particle creation in expanding universes. 1.}",
    doi = "10.1103/PhysRev.183.1057",
    journal = "Phys. Rev.",
    volume = "183",
    pages = "1057--1068",
    year = "1969"
}

@article{PhysRevD.9.341,
  title = {Adiabatic regularization of the energy-momentum tensor of a quantized field in homogeneous spaces},
  author = {Parker, Leonard and Fulling, S. A.},
  journal = {Phys. Rev. D},
  volume = {9},
  issue = {2},
  pages = {341--354},
  numpages = {0},
  year = {1974},
  month = {Jan},
  publisher = {American Physical Society},
  doi = {10.1103/PhysRevD.9.341},
  url = {https://link.aps.org/doi/10.1103/PhysRevD.9.341}
}

@article{Agullo:2018wbf,
    author = "Agullo, Ivan",
    title = "{Primordial power spectrum from the Dapor-Liegener model of loop quantum cosmology}",
    eprint = "1805.11356",
    archivePrefix = "arXiv",
    primaryClass = "gr-qc",
    reportNumber = "LSU-REL-052918",
    doi = "10.1007/s10714-018-2413-1",
    journal = "Gen. Rel. Grav.",
    volume = "50",
    number = "7",
    pages = "91",
    year = "2018"
}

@article{Fernandez-Mendez:2012poe,
    author = "Fernandez-Mendez, Mikel and Mena Marugan, Guillermo A. and Olmedo, Javier",
    title = "{Hybrid quantization of an inflationary universe}",
    eprint = "1205.1917",
    archivePrefix = "arXiv",
    primaryClass = "gr-qc",
    doi = "10.1103/PhysRevD.86.024003",
    journal = "Phys. Rev. D",
    volume = "86",
    pages = "024003",
    year = "2012"
}

@article{Fernandez-Mendez:2013jqa,
    author = "Fern{\'a}ndez-M{\'e}ndez, Mikel and Mena Marug{\'a}n, Guillermo A. and Olmedo, Javier",
    title = "{Hybrid quantization of an inflationary model: The flat case}",
    eprint = "1307.5222",
    archivePrefix = "arXiv",
    primaryClass = "gr-qc",
    doi = "10.1103/PhysRevD.88.044013",
    journal = "Phys. Rev. D",
    volume = "88",
    number = "4",
    pages = "044013",
    year = "2013"
}

@article{Fernandez-Mendez:2014raa,
    author = "Fern{\'a}ndez-M{\'e}ndez, Mikel and Mena Marug{\'a}n, Guillermo A. and Olmedo, Javier",
    title = "{Effective dynamics of scalar perturbations in a flat Friedmann-Robertson-Walker spacetime in Loop Quantum Cosmology}",
    eprint = "1401.5256",
    archivePrefix = "arXiv",
    primaryClass = "gr-qc",
    doi = "10.1103/PhysRevD.89.044041",
    journal = "Phys. Rev. D",
    volume = "89",
    number = "4",
    pages = "044041",
    year = "2014"
}

@article{ElizagaNavascues:2020uyf,
    author = "Elizaga Navascu{\'e}s, Beatriz and Marug{\'a}n, Guillermo A. Mena",
    title = "{Hybrid Loop Quantum Cosmology: An Overview}",
    eprint = "2011.04559",
    archivePrefix = "arXiv",
    primaryClass = "gr-qc",
    doi = "10.3389/fspas.2021.624824",
    journal = "Front. Astron. Space Sci.",
    volume = "8",
    pages = "81",
    year = "2021"
}

@article{Li:2018fco,
    author = "Li, Bao-Fei and Singh, Parampreet and Wang, Anzhong",
    title = "{Qualitative dynamics and inflationary attractors in loop cosmology}",
    eprint = "1807.05236",
    archivePrefix = "arXiv",
    primaryClass = "gr-qc",
    doi = "10.1103/PhysRevD.98.066016",
    journal = "Phys. Rev. D",
    volume = "98",
    number = "6",
    pages = "066016",
    year = "2018"
}

@article{Li:2018opr,
    author = "Li, Bao-Fei and Singh, Parampreet and Wang, Anzhong",
    title = "{Towards Cosmological Dynamics from Loop Quantum Gravity}",
    eprint = "1801.07313",
    archivePrefix = "arXiv",
    primaryClass = "gr-qc",
    doi = "10.1103/PhysRevD.97.084029",
    journal = "Phys. Rev. D",
    volume = "97",
    number = "8",
    pages = "084029",
    year = "2018"
}

@article{Agullo:2012sh,
    author = "Agullo, Ivan and Ashtekar, Abhay and Nelson, William",
    title = "{\em A Quantum Gravity Extension of the Inflationary Scenario}",
    eprint = "1209.1609",
    archivePrefix = "arXiv",
    primaryClass = "gr-qc",
    reportNumber = "IGC-12-9-1",
    doi = "10.1103/PhysRevLett.109.251301",
    journal = "Phys. Rev. Lett.",
    volume = "109",
    pages = "251301",
    year = "2012"
}

@article{Agullo:2013ai,
    author = "Agullo, Ivan and Ashtekar, Abhay and Nelson, William",
    title = "{\em The pre-inflationary dynamics of loop quantum cosmology: Confronting quantum gravity with observations}",
    eprint = "1302.0254",
    archivePrefix = "arXiv",
    primaryClass = "gr-qc",
    doi = "10.1088/0264-9381/30/8/085014",
    journal = "Class. Quant. Grav.",
    volume = "30",
    pages = "085014",
    year = "2013"
}

@article{Agullo:2012fc,
    author = "Agullo, Ivan and Ashtekar, Abhay and Nelson, William",
    title = "{\em Extension of the quantum theory of cosmological perturbations to the Planck era}",
    eprint = "1211.1354",
    archivePrefix = "arXiv",
    primaryClass = "gr-qc",
    reportNumber = "IGC-12-11-3",
    doi = "10.1103/PhysRevD.87.043507",
    journal = "Phys. Rev. D",
    volume = "87",
    number = "4",
    pages = "043507",
    year = "2013"
}

@article{Agullo:2017eyh,
    author = "Agullo, Ivan and Bolliet, Boris and Sreenath, V.",
    title = "{\em Non-Gaussianity in Loop Quantum Cosmology}",
    eprint = "1712.08148",
    archivePrefix = "arXiv",
    primaryClass = "gr-qc",
    doi = "10.1103/PhysRevD.97.066021",
    journal = "Phys. Rev. D",
    volume = "97",
    number = "6",
    pages = "066021",
    year = "2018"
}

@article{Ashtekar:2011ni,
    author = "Ashtekar, Abhay and Singh, Parampreet",
    title = "{\em Loop Quantum Cosmology: A Status Report}",
    eprint = "1108.0893",
    archivePrefix = "arXiv",
    primaryClass = "gr-qc",
    doi = "10.1088/0264-9381/28/21/213001",
    journal = "Class. Quant. Grav.",
    volume = "28",
    pages = "213001",
    year = "2011"
}

@article{deBlas:2016puz,
    author = "de Blas, Daniel Mart\'\i{}n and Olmedo, Javier",
    title = "{\em Primordial power spectra for scalar perturbations in loop quantum cosmology}",
    eprint = "1601.01716",
    archivePrefix = "arXiv",
    primaryClass = "gr-qc",
    doi = "10.1088/1475-7516/2016/06/029",
    journal = "JCAP",
    volume = "06",
    pages = "029",
    year = "2016"
}

@article{Zhu:2017jew,
    author = "Zhu, Tao and Wang, Anzhong and Cleaver, Gerald and Kirsten, Klaus and Sheng, Qin",
    title = "{\em Pre-inflationary universe in loop quantum cosmology}",
    eprint = "1705.07544",
    archivePrefix = "arXiv",
    primaryClass = "gr-qc",
    doi = "10.1103/PhysRevD.96.083520",
    journal = "Phys. Rev. D",
    volume = "96",
    number = "8",
    pages = "083520",
    year = "2017"
}

@article{CastelloGomar:2017kbo,
    author = "Castell\'o Gomar, Laura and Mena Marug\'an, Guillermo A. and Mart\'\i{}n De Blas, Daniel and Olmedo, Javier",
    title = "{\em Hybrid loop quantum cosmology and predictions for the cosmic microwave background}",
    eprint = "1702.06036",
    archivePrefix = "arXiv",
    primaryClass = "gr-qc",
    reportNumber = "IGC-17|2-1",
    doi = "10.1103/PhysRevD.96.103528",
    journal = "Phys. Rev. D",
    volume = "96",
    number = "10",
    pages = "103528",
    year = "2017"
}

@article{Ashtekar:2016wpi,
    author = "Ashtekar, Abhay and Gupt, Brajesh",
    title = "{\em Quantum Gravity in the Sky: Interplay between fundamental theory and observations}",
    eprint = "1608.04228",
    archivePrefix = "arXiv",
    primaryClass = "gr-qc",
    doi = "10.1088/1361-6382/34/1/014002",
    journal = "Class. Quant. Grav.",
    volume = "34",
    number = "1",
    pages = "014002",
    year = "2017"
}

@article{Ashtekar:2016pqn,
    author = "Ashtekar, Abhay and Gupt, Brajesh",
    title = "{\em Initial conditions for cosmological perturbations}",
    eprint = "1610.09424",
    archivePrefix = "arXiv",
    primaryClass = "gr-qc",
    doi = "10.1088/1361-6382/aa52d4",
    journal = "Class. Quant. Grav.",
    volume = "34",
    number = "3",
    pages = "035004",
    year = "2017"
}

@article{ElizagaNavascues:2020fai,
    author = "Elizaga Navascu\'es, Beatriz and Marug\'an, Guillermo A. Mena and Prado, Santiago",
    title = "{\em Non-oscillating power spectra in Loop Quantum Cosmology}",
    eprint = "2005.10194",
    archivePrefix = "arXiv",
    primaryClass = "gr-qc",
    doi = "10.1088/1361-6382/abc6bb",
    journal = "Class. Quant. Grav.",
    volume = "38",
    number = "3",
    pages = "035001",
    year = "2020"
}

@article{Li:2020mfi,
    author = "Li, Bao-Fei and Olmedo, Javier and Singh, Parampreet and Wang, Anzhong",
    title = "{\em Primordial scalar power spectrum from the hybrid approach in loop cosmologies}",
    eprint = "2008.09135",
    archivePrefix = "arXiv",
    primaryClass = "gr-qc",
    doi = "10.1103/PhysRevD.102.126025",
    journal = "Phys. Rev. D",
    volume = "102",
    pages = "126025",
    year = "2020"
}

@article{Li:2019qzr,
    author = "Li, Bao-Fei and Singh, Parampreet and Wang, Anzhong",
    title = "{\em Primordial power spectrum from the dressed metric approach in loop cosmologies}",
    eprint = "1912.08225",
    archivePrefix = "arXiv",
    primaryClass = "gr-qc",
    doi = "10.1103/PhysRevD.101.086004",
    journal = "Phys. Rev. D",
    volume = "101",
    number = "8",
    pages = "086004",
    year = "2020"
}

@article{Yang:2009fp,
    author = "Yang, Jinsong and Ding, You and Ma, Yongge",
    title = "{\em Alternative quantization of the Hamiltonian in loop quantum cosmology II: Including the Lorentz term}",
    eprint = "0904.4379",
    archivePrefix = "arXiv",
    primaryClass = "gr-qc",
    reportNumber = "AEI-2009-042",
    doi = "10.1016/j.physletb.2009.10.072",
    journal = "Phys. Lett. B",
    volume = "682",
    pages = "1--7",
    year = "2009"
}

@article{Li:2022evi,
    author = "Li, Bao-Fei and Singh, Parampreet",
    title = "{\em Close relationship between the dressed metric and the hybrid approach to perturbations in effective loop quantum cosmology}",
    eprint = "2206.12434",
    archivePrefix = "arXiv",
    primaryClass = "gr-qc",
    doi = "10.1103/PhysRevD.106.086015",
    journal = "Phys. Rev. D",
    volume = "106",
    number = "8",
    pages = "086015",
    year = "2022"
}

@article{Iteanu:2022zha,
    author = "Iteanu, Simon and Mena Marug\'an, Guillermo A.",
    title = "{\em Mass of Cosmological Perturbations in the Hybrid and Dressed Metric Formalisms of Loop Quantum Cosmology for the Starobinsky and Exponential Potentials}",
    eprint = "2208.01987",
    archivePrefix = "arXiv",
    primaryClass = "gr-qc",
    doi = "10.3390/universe8090463",
    journal = "Universe",
    volume = "8",
    number = "9",
    pages = "463",
    year = "2022"
}

@article{Li:2019ipm,
    author = "Li, Bao-Fei and Singh, Parampreet and Wang, Anzhong",
    title = "{\em Genericness of pre-inflationary dynamics and probability of the desired slow-roll inflation in modified loop quantum cosmologies}",
    eprint = "1906.01001",
    archivePrefix = "arXiv",
    primaryClass = "gr-qc",
    doi = "10.1103/PhysRevD.100.063513",
    journal = "Phys. Rev. D",
    volume = "100",
    number = "6",
    pages = "063513",
    year = "2019"
}

@article{Planck:2018jri,
    author = "Akrami, Y. and others",
    collaboration = "Planck",
    title = "{Planck 2018 results. X. Constraints on inflation}",
    eprint = "1807.06211",
    archivePrefix = "arXiv",
    primaryClass = "astro-ph.CO",
    doi = "10.1051/0004-6361/201833887",
    journal = "Astron. Astrophys.",
    volume = "641",
    pages = "A10",
    year = "2020"
}

@article{Saeed:2024xhk,
    author = "Saeed, Jamal and Pan, Rui and Brown, Christian and Cleaver, Gerald and Clevear, Gerald and Wang, Anzhong",
    title = "{Universal Properties of the Evolution of the Universe in Modified Loop Quantum Cosmology}",
    eprint = "2406.06745",
    archivePrefix = "arXiv",
    primaryClass = "gr-qc",
    doi = "10.3390/universe10100397",
    journal = "Universe",
    volume = "10",
    number = "10",
    pages = "397",
    year = "2024"
}

@article{Agullo:2023rqq,
    author = "Agull\'o, Ivan and Wang, Anzhong and Wilson-Ewing, Edward",
    title = "{Loop quantum cosmology: relation between theory and observations}",
    eprint = "2301.10215",
    archivePrefix = "arXiv",
    primaryClass = "gr-qc",
    doi = "10.1007/978-981-19-3079-9_103-1",
    month = "1",
    year = "2023"
}

@article{Penna-Lima:2022dmx,
    author = "Penna-Lima, Mariana and Pinto-Neto, Nelson and Vitenti, Sandro D. P.",
    title = "{New formalism to define vacuum states for scalar fields in curved spacetimes}",
    eprint = "2207.08270",
    archivePrefix = "arXiv",
    primaryClass = "gr-qc",
    doi = "10.1103/PhysRevD.107.065019",
    journal = "Phys. Rev. D",
    volume = "107",
    number = "6",
    pages = "065019",
    year = "2023"
}

@article{Zhu:2013upa,
    author = "Zhu, Tao and Wang, Anzhong and Cleaver, Gerald and Kirsten, Klaus and Sheng, Qin",
    title = "{Inflationary cosmology with nonlinear dispersion relations}",
    eprint = "1308.5708",
    archivePrefix = "arXiv",
    primaryClass = "astro-ph.CO",
    doi = "10.1103/PhysRevD.89.043507",
    journal = "Phys. Rev. D",
    volume = "89",
    number = "4",
    pages = "043507",
    year = "2014"
}

@article{Zhu:2014aea,
    author = "Zhu, Tao and Wang, Anzhong and Cleaver, Gerald and Kirsten, Klaus and Sheng, Qin",
    title = "{Gravitational quantum effects on power spectra and spectral indices with higher-order corrections}",
    eprint = "1405.5301",
    archivePrefix = "arXiv",
    primaryClass = "astro-ph.CO",
    doi = "10.1103/PhysRevD.90.063503",
    journal = "Phys. Rev. D",
    volume = "90",
    number = "6",
    pages = "063503",
    year = "2014"
}

@article{Zhu:2016srz,
    author = "Zhu, Tao and Wang, Anzhong and Kirsten, Klaus and Cleaver, Gerald and Sheng, Qin",
    title = "{High-order Primordial Perturbations with Quantum Gravitational Effects}",
    eprint = "1604.05739",
    archivePrefix = "arXiv",
    primaryClass = "gr-qc",
    doi = "10.1103/PhysRevD.93.123525",
    journal = "Phys. Rev. D",
    volume = "93",
    number = "12",
    pages = "123525",
    year = "2016"
}

@article{Pan:2023aiv,
    author = "Pan, Rui and Marchetta, John Joseph and Saeed, Jamal and Cleaver, Gerald and Li, Bao-Fei and Wang, Anzhong and Zhu, Tao",
    title = {{Uniform Asymptotic Approximation Method with P\"oschl\textendash{}Teller Potential}},
    eprint = "2309.03327",
    archivePrefix = "arXiv",
    primaryClass = "gr-qc",
    doi = "10.3390/universe9110471",
    journal = "Universe",
    volume = "9",
    number = "11",
    pages = "471",
    year = "2023"
}

@inbook{Olver:1997, place={Wellesley, Massachusetts}, title={Asymptotics and Special Functions},  publisher={A K Peters, Ltd.}, author={Olver, Frank W. J.}, year={1997}, pages={1–572}}

@article{AgulloMorris2015,
    author = "Agullo, Ivan and Morris, Noah A.",
    title = "{Detailed analysis of the predictions of loop quantum cosmology for the primordial power spectra}",
    doi = "10.1103/PhysRevD.92.124040",
    journal = "Phys. Rev. D",
    volume = "92",
    number = "12",
    pages = "124040",
    year = "2015"
}

@article{AgulloOlmedoSreenath2020PRL,
    author = "Agullo, Ivan and Olmedo, Javier and Sreenath, V.",
    title = "{Predictions for the Cosmic Microwave Background from an Anisotropic Quantum Bounce}",
    doi = "10.1103/PhysRevLett.124.251301",
    journal = "Phys. Rev. Lett.",
    volume = "124",
    number = "25",
    pages = "251301",
    year = "2020"
}

@article{AgulloOlmedoSreenath2020PRD,
    author = "Agullo, Ivan and Olmedo, Javier and Sreenath, V.",
    title = "{Observational consequences of Bianchi I spacetimes in loop quantum cosmology}",
    doi = "10.1103/PhysRevD.102.043523",
    journal = "Phys. Rev. D",
    volume = "102",
    number = "4",
    pages = "043523",
    year = "2020"
}

@article{AgulloNelsonAshtekar2015,
    author = "Agullo, Ivan and Nelson, William and Ashtekar, Abhay",
    title = "{Preferred instantaneous vacuum for linear scalar fields in cosmological space-times}",
    doi = "10.1103/PhysRevD.91.064051",
    journal = "Phys. Rev. D",
    volume = "91",
    number = "6",
    pages = "064051",
    year = "2015"
}

@article{AgulloNavarroSalasOlmoParker2010,
    author = "Agullo, Ivan and Navarro-Salas, Jose and Olmo, Gonzalo J. and Parker, Leonard",
    title = "{Revising the observable consequences of slow-roll inflation}",
    doi = "10.1103/PhysRevD.81.043514",
    journal = "Phys. Rev. D",
    volume = "81",
    number = "4",
    pages = "043514",
    year = "2010"
}
\bibliographystyle{h-physrev}
\end{document}